\documentclass{aastex61}
\usepackage{makecell}

\shorttitle{GRB prompt emission polarimetry with CZTI}
\shortauthors{Chattopadhyay et al.}

\begin{document}

\title{Prompt emission polarimetry of Gamma Ray Bursts with {\em AstroSat} CZT-Imager}

\correspondingauthor{Tanmoy Chattopadhyay}
\email{tanmoyc@stanford.edu}

\author{Tanmoy Chattopadhyay}
\affil{Pennsylvania State University,
University Park, PA 16802, USA}
\affil{Physical Research Laboratory, 
Ahmedabad, Gujarat, India}
\affil{Department of Physics, Stanford University, 382 Via Pueblo Mall, Stanford CA 94305, USA}
\affil{ Kavli Institute of Astrophysics and Cosmology, 452 Lomita Mall, Stanford, CA 94305, USA}

\author{Santosh V. Vadawale}
\affil{Physical Research Laboratory,
Ahmedabad, Gujarat, India}

\author{E. Aarthy}
\affil{Physical Research Laboratory,
Ahmedabad, Gujarat, India}

\author{N. P. S. Mithun}
\affil{Physical Research Laboratory,
Ahmedabad, Gujarat, India}

\author{Vikas Chand}
\affiliation{Tata Institute of Fundamental Research,
Mumbai, India}

\author{Ajay Ratheesh}
\affiliation{Tata Institute of Fundamental Research,
Mumbai, India}
\affiliation{Dipartimento di Fisica, Università di Roma Tor Vergata, Via della Ricerca Scientifica 1, I-00133 Roma, Italy}
\affiliation{INAF Istituto di Astrofisica e Planetologia Spaziali, Via del Fosso del Cavaliere 100,
00133 Roma (RM), Italy}
\affiliation{Dipartimento di Fisica, Università La Sapienza, P. le A. Moro 2, 00185 Roma, Italy}

\author{Rupal Basak}
\affiliation{INAF--IASF Bologna, via P. Gobetti, 101, 40129 Bologna, Italy}
\affiliation{Dipartimento di Fisica e Scienze della Terra, Universit\'a di Ferrara, via Saragat 1, I-44122, Ferrara, Italy}
\affiliation{Department of Physics, KTH Royal Institute of Technology, and the Oskar Klein Centre for Cosmoparticle Physics,
10691 Stockholm, Sweden}

\author{A. R. Rao}
\affiliation{Tata Institute of Fundamental Research,
Mumbai, India}

\author{Varun Bhalerao}
\affiliation{Indian Institute of Technology Bombay, Mumbai, India}

\author{Sujay Mate}
\affiliation{Indian Institute of Technology Bombay, Mumbai, India}
\affiliation{IRAP, Universit\'e de Toulouse, CNES, CNRS, UPS, Toulouse, France}

\author{Arvind B.}
\affiliation{Indian Institute of Technology Bombay, Mumbai, India}
\affiliation{Physics and Astronomy Department, Texas Tech University, Lubbock, USA}

\author{V. Sharma}
\affiliation{The Inter-University Centre for Astronomy and Astrophysics,
Pune, India}

\author{Dipankar Bhattacharya}
\affiliation{The Inter-University Centre for Astronomy and Astrophysics,
Pune, India}

\begin{abstract}
X-ray and Gamma-ray polarization measurements 
of the prompt emission of Gamma-ray bursts (GRBs) are 
believed to be extremely important for testing various models of GRBs. So far, the available measurements of hard X-ray  polarization 
of GRB prompt emission have not significantly constrained
the GRB models, particularly because of the difficulty of 
measuring polarization in these bands. 
The CZT Imager (CZTI) onboard {\em AstroSat} is primarily an X-ray spectroscopic 
instrument that also works as a wide angle GRB monitor due to the transparency 
of its support structure above 100 keV. It also has experimentally verified 
polarization measurement capability in the 100 $-$ 300 keV energy range and 
thus provides a unique opportunity to attempt spectro-polarimetric studies of GRBs.
Here we present the polarization data for the brightest 11 GRBs detected by CZTI  
during its first year of operation.
Among these, 5 GRBs show polarization signatures with
$\gtrapprox$3$\sigma$, and 1 GRB shows $\>$2$\sigma$ 
detection significance.  
We place upper limits for the remaining 5 GRBs. 
We provide details of the various tests performed to validate our
polarization measurements. 
While it is difficult yet to discriminate between various emission models 
with the current sample alone, the large number of polarization measurements 
CZTI expects to gather in its minimum lifetime of five years should help to
significantly improve our understanding of the prompt emission.       

\end{abstract}


\section{Introduction} \label{sec1}

Gamma-ray bursts (GRBs) are thought to accompany the birth of stellar mass black holes in 
the core collapse of massive stars 
\citep{woosley93, iwamoto98,macfadyen99} or mergers of compact star
 binaries \citep{eichler89,narayan92}. Phenomenologically, GRB 
emission occurs in two distinct phases -- the prompt and the afterglow.
The initial burst of high energy emission or the prompt emission is widely 
believed to originate from a jet close to the black hole, while the 
long-lasting multi-wavelength afterglow emission is generated far from the compact object by the interaction of the GRB 
jet with the ambient medium \citep{piran05,meszaros06}.  
Despite the observation of a large number of GRBs in the 
last decade \citep{gehrels12} with sensitive detectors on board {\em Swift} \citep{gehrels04,barthelmy05} and {\em Fermi} \citep{meegan09} missions, the mechanism of the prompt emission has not yet been well 
understood \citep{kumar15} owing to the diversity, extreme variability and very short duration (seconds) of this phase
\citep{hakkila14,basak17}.
The prompt emission is believed to be generated either via the  synchrotron process \citep{rees94,sari98} or via 
inverse Compton scattering 
\citep{ghisellini99,ghisellini00,lazzati04}. In addition, a few cases display the evidence of a thermal blackbody component, presumably of
photospheric origin \citep{ryde04,peer11,basak15,iyyani15}. One way of distinguishing between these emission processes would be through their
unique polarization signatures, so the measurement of 
X-ray and Gamma ray polarization is considered to be of great importance in the study of GRB prompt emission \citep{covino16,mcconnell16}. In particular, a statistical study of GRB polarization could provide critical constraints on the geometry and the radiation process involved \citep{toma08}.

The past decade has seen several attempts to measure the X-ray/gamma-ray polarization of GRBs. Instruments such as RHESSI \citep{coburn03}, IBIS \citep{gotz13,gotz14} and 
SPI \citep{mcglynn07,kalemci07,mcglynn09} onboard {\em INTEGRAL}, as well as
BATSE \citep{willis05} onboard {\em CGRO} \citep[see a review by][] {mcconnell16} have reported several cases of strong polarization. However, these instruments being of primarily spectroscopic nature and not optimized for polarimetry, the results, limited by statistical and systematic uncertainties, are often thought to be unreliable \citep{rutledge04,wigger04}. 
Subsequently GAP \citep{yonetoku06}, a dedicated large FOV Compton polarimeter flown 
in 2011, obtained polarization measurements for three bright GRBs 
\citep{yonetoku11,yonetoku12}. More recently, 
POLAR \citep{sun16,orsi11}, a dedicated GRB polarimeter launched in 2016, provided precise polarization measurements for five GRBs in their prompt phase in hard X-rays \citep{zhang19}. They found the polarization of the GRBs to be low, between 4 and 11\%, hinting at the unpolarized nature of GRBs in general. However, for firm confirmation, a larger sample with such precise measurements is necessary.  POLAR, however,  stopped its operation on 2017 March  31.

{\em AstroSat} \citep{singh14}, India's first dedicated astronomical satellite, was launched on 2015 September 28, and has been operating successfully. 
The Cadmium Zinc Telluride Imager (CZTI) instrument
onboard {\em AstroSat}, with an array of CZT 
detectors, is a large area ($\sim$1000 cm$^2$) spectroscopic 
instrument with a coded mask imaging capability in the energy range of
20 -- 150 keV \citep{bhalerao16,vadawale16,chattopadhyay16}.
The 5 mm thick pixelated CZT detectors used in this instrument present significant Compton scattering probability in hard X-rays, thereby enabling it to operate as a Compton polarimeter in the energy range 100$-$300 keV, as demonstrated during ground calibration
\citep{chattopadhyay14,vadawale15}. At these energies, the supporting 
structure of CZTI becomes increasingly transparent, making the instrument capable of detecting transient events like GRBs and performing their polarimetric measurements.
On the very first day of its operation, CZTI detected a
GRB  \citep[GRB 151006A,][]{bhalerao15} at an angle 60$^\circ$ from the pointing direction. A detailed spectro-polarimetric study of GRB 151006A has been reported in \citet{rao16}.

Apart from GAP and POLAR, the previously reported polarization 
results by {\em RHESSI} and {\em INTEGRAL}
have the drawback that the instruments were never calibrated 
before flight with polarized
and unpolarized sources, which draws a lot of criticism regarding the reliability
of the reported results. CZTI, on the other hand, has been extensively
calibrated for polarization measurements before launch, thus boosting
the confidence in the obtained results. Additionally, because of the 
large collecting area of the instrument and the favorable
Compton scattering geometry presented by the pixels, the polarimetric 
sensitivity of CZTI is significantly higher than other contemporary X-ray polarimeters.
Given a minimum lifetime of 5 years of 
{\em AstroSat}, we expect to obtain a large sample of GRB polarization
with CZTI which, along with results from other existing/upcoming GRB polarimetry 
missions, may lead to a better understanding of these objects.     

CZTI in the first one year of operation (2015 October 6 
to 2016 October 5) detected a total of  47 GRBs, among which we
attempted polarization measurements for the 11 brightest events. We
select GRBs with fluence higher than 10$^{-5}$ erg cm$^{-2}$ so that
the number of Compton events are sufficient to attempt polarization
measurements. Most of these GRBs appear to show signatures of high polarization. 
In this paper we report the detailed data analysis and the resulting polarization 
estimates for these GRBs. These are, however, difficult measurements due to
the scarcity of flux in most cases and the extreme photon hungry nature of
X-ray polarimetry. 
We have treated the statistical uncertainties and the possible sources
of systematics which may introduce false polarimetric signature, 
with utmost care for each of the GRBs.
In section  \ref{sec2}, we discuss the polarization capability of  
CZTI, the analysis procedure, and the details of the individual GRBs in 
our sample. This is followed by the final results and discussions
in sections \ref{sec3} and \ref{sec4} respectively. 

\section{CZTI as a GRB polarimeter: Analysis procedure} \label{sec2}
CZTI consists of an array of 64 CZT modules where each detector is 
5 mm thick, providing high quantum efficiency and fine spectral
resolution in a broad energy band from a few keV to a few hundred keV. Each
detector module is further pixelated into 256 pixels (with a nominal 
pixel size of 2.5 mm $\times$ 2.5 mm). A 0.5 mm thick
Tantalum coded mask provides imaging capability to the instrument in 20--150 keV 
energy range.
CZTI also has the advantage of working in a photon tagging mode with a 
time resolution of 20 $\mu$s. All these features make CZTI 
well suited to study the spectral and timing features of
celestial X-ray objects in the 20--150 keV region. 
   
Besides the spectroscopic and timing studies, CZTI also 
works as a sensitive Compton polarimeter for bright X-ray sources
at higher energies. This feature arises out of the significant 
Compton scattering cross-section of 5 mm thick CZT detectors at energies 
beyond 100 keV and the availability of continuous time tagged events 
from CZTI. Compton scattering occurs preferentially
in a direction perpendicular to the incident polarization, giving
rise to a sinusoidal modulation in the distribution of 
azimuthal scattering angles \citep{lei97,kaaret14}. The flight 
configuration of CZTI has been shown to possess polarization measurement capability,
through detailed experiments and simulation studies during its ground calibration
\citep{chattopadhyay14,vadawale15}. 
Because of the increasing transparency of the collimators and 
the supporting structure in the 100--300 keV range, CZTI  
works as an open detector and captures high energy transient events like 
GRBs occurring all over the sky. Detection of GRB 151006A on the very first 
day of its operation \citep{bhalerao15}, at a large off-axis angle, demonstrates 
the capability of CZTI as a wide angle GRB monitor.  This, coupled with the polarimeric
capability of CZTI, offers an unique opportunity to attempt polarization measurements of 
GRBs with CZTI, particularly since GRB prompt emission could be strongly polarized. 
GRB polarimetry with CZTI is very similar to the On-axis polarimetry of persistent sources, but with the following key advantages.
\begin{itemize}
\item Because CZTI polarimetric observations do not require any change 
in the hardware configuration, polarimetric analysis can be attempted 
from data obtained in the standard mode. CZTI detects 4--5 GRBs in a 
month. Polarimetric analysis can in principle be attempted for any 
detected GRB. 
\item GRB prompt emission is expected to be strongly polarized 
owing to its non-thermal origin and the involvement of high bulk lorentz factors, thus 
making detection easier.
\item Compared to bright persistent sources like Crab or Cygnus X-1, GRBs 
provide higher signal to noise ratio in Compton events resulting in 
a higher polarimetric sensitivity.
\item Accurate polarimetric background measurements are available just
before and after the GRB event. 
\end{itemize}

\subsection{Criteria for the selection of Compton events}\label{subsec2}
The selection criteria of the Compton events has been discussed in detail in 
\citet{chattopadhyay14}. Each event in the CZTI output data has an 
individual time stamp with a resolution of 20 $\mu$s. Any two events
occurring within 20 $\mu$s will bear the same time stamp. The event file also
lists the pixel and detector ID, the PHA channel of detection, veto
and alpha coincidence flags.
The Compton scattered events are normally expected to be
captured within a 20 $\mu$s time window. However, since the readout in CZTI is 
done for one module at a time, if two events are registered in two 
different pixels in the same module,
there is a certain probability that the two events would get different 
time stamps. Therefore, we select all the double pixel events happening within a
coincidence window of 40 $\mu$s, as polarization information of the 
radiation is embedded in these double pixel events. Events at three, four or
more pixels within a coincidence window are excluded 
from analysis, primarily because the probability of such events due to Compton 
scattering is low and it is difficult to identify the first event out of these
multiple events. In case of double pixel events, the pixel with the lower energy
deposition is considered to be the scattering pixel and the higher energy pixel
as the absorbing pixel. It is to be noted that CZTI detectors provide better than 10 \% and 5 \% energy resolution at 59.5 and 122 keV respectively. In Compton scattering, the electron recoil energy (lower energy deposition) is normally well below the scattered photon energy and therefore they are widely separated and easily distinguishable. A time window as large as 40 $\mu$s may also 
result in false chance coincidence events. These events are filtered
out by applying the Compton kinematics criteria: 1) spatial proximity 
of the pixels and 2) sum and ratio of the deposited energies must be
consistent with those expected for true Compton events for the scattering
geometry of CZTI.
We also exclude all the veto and alpha flagged events (``alpha flag" refers to simultaneous detection of an alpha particle by a CsI scintillator kept inside the Am$^{241}$ onboard calibration box and a 59.5 keV X-ray photon in the CZT detector) from further 
analysis as these events do not carry any polarization information 
and therefore contribute indirectly to the background.

\begin{figure}
\centering
\includegraphics[scale=.7]{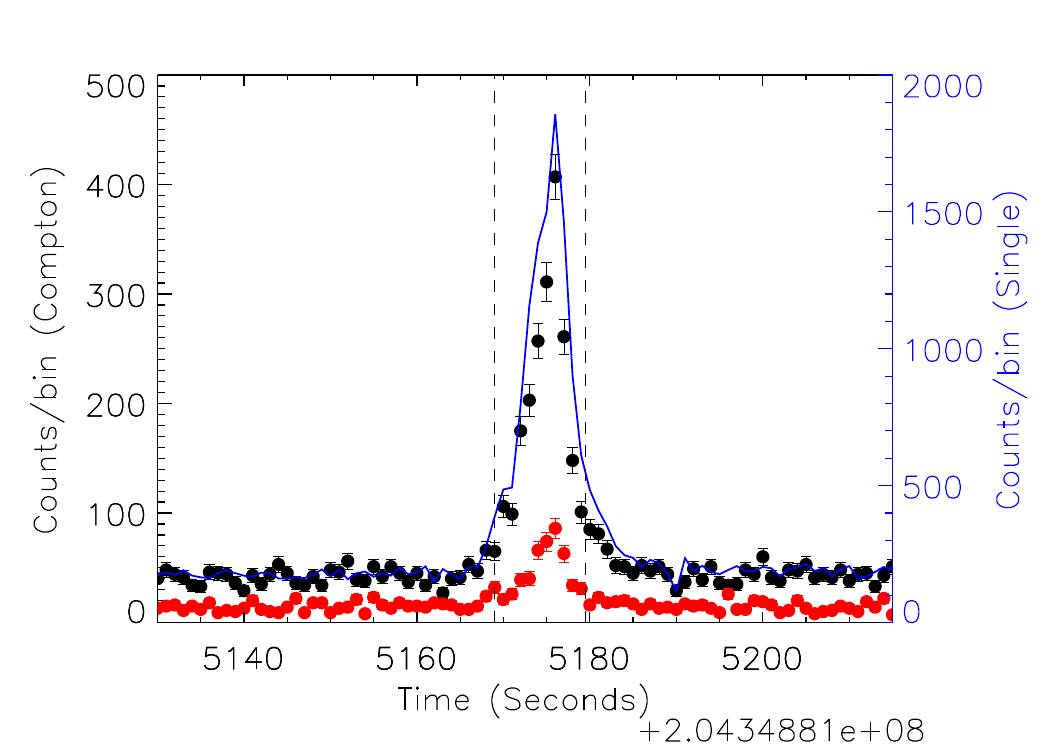}
\caption{Observed rate of single and double events in CZTI during GRB 160623A. 
The blue solid line (plotted against the right axis) is obtained
from the detected single events. 
The events satisfying the Compton criteria (plotted against the left axis) 
are shown in 
black and the red data points (plotted against the left axis) are double
events not satisfying the 
Compton criteria. The region between the dashed vertical lines 
in the light curve marks the prompt emission phase of GRB 160623A. 
The Compton events within this region are used for further analysis.}
\label{fig_lc}
\end{figure}

Figure \ref{fig_lc} shows the light curve of GRB 160623A in single (blue line) and Compton events (black data points). 
A clear detection of the GRB in Compton events shows the pertinence of 
the event selection criteria.
It is to be noted that CZTI has observed the Crab nebula for $\sim$790 ks 
and could obtain a clear pulse profile of Crab pulsar in the Compton 
events in the 100--380 keV range (for On-axis sources, Compton events 
are selected in 100--380 keV energy range)  
which also independently validates the Compton event selection 
algorithm \citep{vadawale17}. In order to further make sure that the 
peak in the Compton 
events is not a result of chance coincidence of the GRB photons, we 
generate a light curve for double pixel events not satisfying the Compton 
conditions. As we see in Figure \ref{fig_lc} (red data points), the GRB does not 
show up as clearly as in case of Compton events. The
small peak in the non-Compton events arises due to a significant 
probability of chance coincidence, given the high flux of the GRB. This is 
expected to be more prominent for brighter GRBs. We discuss about the 
estimation of these events and their effect on the polarization analysis
in a later section.

\subsection{The GRB sample for our study}\label{subsec1}
In the first one year of operation of CZTI
(2015 October 6 to  2016 September 30) a total of 47 GRBs were detected\footnote{\url{http://astrosat.iucaa.in/czti/?q=grb}}. 
Out of the 47 GRBs, we selected 11 GRBs which are bright 
enough to give sufficient number of Compton events
(the number of double events satisfying the Compton criterion greater than
400) to attempt  polarization analysis.
Out of these 47 GRBs, 33 were detected by {\em Fermi}/GBM, 14 by
{\em Swift}/BAT, with 8 being common between both these 
instruments. Localization of these GRBs in CZTI co-ordinates was done using
the position information available in the {\em Swift} and {\em Fermi} GRB data 
bases. Our choice of bright GRBs with sufficient Compton events corresponds to 
a limiting fluence of $10^{-5}$~erg~cm$^{-2}$.  The {\em Fermi} and {\em Swift} 
catalogues list 18 GRBs above this limit during the corresponding 1 year period.
{\em AstroSat} has $\sim 30$\% of the sky occulted by the Earth at any given time, 
and during this year had data gaps for $\sim 20.5$\% of the time due to passage through 
the South Atlantic Anomaly (SAA) or telemetry errors. The detection of 
11 of the 18 bright GRBs in CZTI ($\sim 60$\%) is consistent with its effective
duty cycle and sky coverage. 

The observed properties of these 11 GRBs are 
listed in Table \ref{table1}. 
Seven of these have triggered {\em Fermi}/GBM detectors 
(listed chronologically in the table), 5 have triggered {\em Swift}/BAT 
detectors, while two of them triggered both these detectors. 
The three GRBs triggered only in {\em Swift}/BAT are listed next in the table 
(again chronologically). GRB 160623A, listed last in the table, is a long GRB 
that triggered Konus/Wind, but was occulted by the Earth for a large part for 
{\em Fermi}/GBM. It however has localization information 
from {\em Swift}/XRT and prompt spectral information from CZTI detectors. 
The GRB location error circles quoted in the table are based on those
provided by the {\em Swift} and {\em Fermi} satellites: the localization
information is taken, as per availability, from {\em Swift}/XRT, {\em Swift}/BAT 
and {\em Fermi}/GBM catalogs.  
The durations of the GRBs, $T_{90}$, are measured in 50--300 keV 
and 15--150 keV bands for the GBM and the BAT detections respectively and are 
collected from various GCN circulars as well as the respective websites. 
For GRB 160623A, however,  $T_{90}$ is obtained from CZTI-Veto. The time durations 
selected for polarization analysis ($t_1$ and $t_2$) are given with respect to the 
trigger time of {\em Fermi}/GBM (for the first 7 GRBs),
{\em Swift}/BAT (for the next 3 GRBs), and CZTI (for GRB 160623A). 
There are multiple values of the duration for 3 GRBs, determined based on 
the availability of the Compton events in the CZTI data.
If afterglow measurements exist, they are indicated in the table with 
symbols X ({\em Swift}/ XRT X-rays),
O (optical), U ({\em Swift}/UVOT), R (radio), and NIR (near-Infrared): 
this information is gathered from available GCN circulars on these GRBs.

In order to compare the observed polarization fractions with theoretical 
predictions, estimates of the peak energy of the GRBs are required \citep{toma08}. 
To obtain the peak energies of the selected GRBs, we carried out a spectroscopic
analysis using the data obtained from GBM, BAT, and CZTI. The GBM and the LAT
low-energy events data were retrieved from the {\em Fermi} Science Support Center
archives\footnote{\url{https://fermi.gsfc.nasa.gov/ssc/data/access/}}. 
The spectral analysis was done for the same time intervals
that have been used for polarization measurements.
The photon spectra were fitted with a Band model \citep{band93} for the 7 GRBs 
detected by GBM, and with a powerlaw with an exponential
cut-off ($\propto E^{-p}exp(-E/E_c)$) for the 3 GRBs detected by BAT.
From the spectral
parameters, we calculated the fluence in the selected time intervals and
energy range 100--300 keV as well as the time integrated fluence in the 
10--1000 keV band (given in parentheses in the last column of the table).
For the three BAT-detected GRBs, we use the Konus/Wind spectral parameters (given in 
the second line in the table) to determine the 10--1000 keV time integrated fluence.
For GRB 160623A, we use Konus/Wind data to determine the spectral parameters 
using the Band model which are then combined with the 
CZTI-Veto spectrum to determine the fluence.
The relevant model parameters are provided in Table \ref{table1}.
The peak energies are in near agreement with the time-integrated
peak energies given in the respective catalogs or GCN circulars. 
The errors in the parameters quoted here represent 90\% confidence 
intervals.
\begin{table*}%
\centering
\begin{tiny}
\caption{The sample of GRBs selected for polarization study with CZTI}

%\hspace{-0.85in}
\begin{tabular}{p{1.1cm}ccp{1.3cm}cccp{1.3 cm}cp{1.2 cm}p{1.2cm}}
\\
\hline\\[0.03cm]
GRB  &Localization${}^{a}$& $T_{90}$& ($t_1$, $t_2$)${}^{b}$ & &Spectral Parameters${}$   &   &  Afterglows${}^{c}$ &Fluence${}^{d}$ & Incident Direction ($\theta, \phi$) & Parts Passing Through\\
(Detectors) &   & (s)  &       (s)        & $\alpha$/$-p$       &   $\beta$         & $E_p$/ $E_c$   &               & &                  \\ 
             &     &               &                   &               &   & (keV)           &   &       &  ($^\circ$)  \\ \hline \\[0.03cm]
151006A (GBM) &$2^{\arcsec}.3$& 84.0& (0,37.0) & $-1.17_{-0.07}^{+0.08}$& $-2.16_{-0.18}^{+0.05}$& $447_{-121}^{+207}$&      X  &0.5(1.6) & 60.82, 67.57 &UVIT, SXT, CZTI collimator  \\[0.03cm] \\
160106A (GBM) &$1^{\arcdeg}.1$&39.4& (-1.5,14.7)& $-0.53_{-0.06}^{+0.07}$& $-2.31_{-0.21}^{+0.14}$& $400_{-40}^{+45}$&      — &3.5(5.6) &  106.12, 255.69 & CZTI collimator\\[0.03cm] \\
160325A (GBM, BAT) &$1^{\arcsec}.7$&43.0& (-0.8,15.2) (39.2,47.2)& $-0.71_{-0.06}^{+0.07}$& $-2.26_{-0.30}^{+0.20}$& $238_{-22}^{+25}$&     X, U, O, NIR &0.76(4.78) & 0.66, 159.44 & Coded mask \\
160509A (GBM)&$2^{\arcsec}.3$ &371.0& (3.7,20.6) & $-0.75_{-0.02}^{+0.02}$& $-2.13_{-0.03}^{+0.03}$& $334_{-10}^{+12}$&      X, O, R  &4.5(48.7) &105.74, 85.45 & CZTI collimator, LAXPC\\[0.03cm]\\
160802A (GBM)&$1^{\arcdeg}.0$ &16.4 & (-1.0,4.0) (12.0,19.0)  &$-0.61_{-0.04}^{+0.04}$& $-2.40_{-0.13}^{+0.10}$& $280_{-14}^{+17}$&          —  &2.2(8.8) & 52.96, 273.12 & CZTI collimator\\
160821A (GBM, BAT) &$1^{\arcmin}.0$&43.0& (130,149)& $-0.97_{-0.01}^{+0.01}$& $-2.25_{-0.03}^{+0.03}$& $866_{-24}^{+25}$&            O  &20.0(47.0) & 156.18, 59.31& Satellite body (below)\\[0.03cm]\\
160910A (GBM)&$4.^{\arcsec}3$ &24.3& (5.9,10.4) &$-0.36_{-0.03}^{+0.03}$& $-2.38_{-0.06}^{+0.05}$& $330_{-13}^{+13}$&         X, O, R  & 0.42(12.3) & 65.54, 333.45 & LAXPC, CZTI collimator\\[0.03cm]\\ \hline \\[0.03cm] 
160131A (BAT)&	$2.^{\arcsec}2$ &325.0& (26.2,42.4) &$-1.00_{-0.14}^{+0.14}$&         ---                             & $388_{-185}^{+2735}$  &                 X, U, O, R  &0.9(6.8) & 116.86, 184.64 & Radiator plate, CZTI collimator\\[0.03cm] \\ 
             &                  &     &             & $-1.16_{-0.04}^{+0.04}$&         $-1.56_{-0.10}^{+0.07}$                             & $586_{-259}^{+518}$  &            &      & \\[0.03cm] \\
160607A (BAT) &$1.^{\arcsec}5$&33.4& (3.3,16)& $-0.9_{-0.1}^{+0.1}$&         ---         &$131_{-24}^{+36}$&                  X, O  &  0.8(3.9) & 138.85, 315.78&Solar panel, satellite body \\[0.03cm]\\
             &                  &     &             & $-1.11_{-0.04}^{+0.04}$&         $-2.50_{-0.35}^{+0.26}$                             & $176_{-42}^{+25}$  &                 &  & \\[0.03cm] \\
160703A (BAT)&$3.^{\arcsec}9$&44.4&  (-4.2,2.9) (3.8,24.2)&$-0.97_{-0.14}^{+0.14}$&         ---             &$277_{-107}^{+430}$    &                        X, U, O, R  &0.6(1.6) & 10.14, 95.05 & Coded mask, CZTI collimator\\[0.03cm]
             &                  &     &             & $-1.23_{-0.04}^{+0.04}$&         ---                            & $327_{-36}^{+46}$  &                 &  & \\[0.03cm] \\ \hline \\[0.03cm]
160623A (CZTI)      &$3.^{\arcsec}5$& 90.4 & (0,7) &$-0.88_{-0.05}^{+0.05}$ &   $-2.95_{-0.14}^{+0.11}$       & $648_{-32}^{+33}$     &                  X, O, NIR, R   & 5.3(18.0) & 140.46, 118.06 & Satellite body (below)\\[0.03cm]\\
\hline
\end{tabular}
%\end{tiny}
$^a$Localization given with 90 \% error radius, taken from Swift/XRT, Swift/BAT, and {\em Fermi}/GBM catalogs. For GRB 160910A, the error 
is only statistical.\\
$^b$t$_1$ ~ and ~ t$_2$ ~are ~w.r.t.~ GBM/BAT~ trigger-time; 
For GRB 160623A  w.r.t. CZTI trigger time: UT 204353981.02834 (seconds since Jan 1, 2010 00:00:00 UTC)\\
Konus/Wind observations of GRB 160131A, GRB 160607A, GRB 160703A and GRB 160623A: \citet{tsvetkova16_grb160131A} (GCN 18974), \citet{tsvetkova16_160607A} (GCN 19511), \citet{frederiks16_160703A} (GCN 19649) and \citet{frederiks16_160623A} (GCN 19554) \\
$^c$Afterglows — O: optical, X: X-rays, R: radio, NIR: near infra-red and U: UVOT\\
$^{d}$Fluence in units of $10^{-5} ~ergs~cm^{-2}$ in the range $t_1$ to $t_2$; 100 $-$ 300 keV; Values in bracket are fluence in 10 $-$ 1000 keV band for the time integrated GRB.\\
\label{table1}
\end{tiny}
\end{table*}
Except for GRB 160325A, all the GRBs were detected outside of CZTI's primary FOV.  In  
such off-axis detections, the radiation passes through 
different parts of the satellite structure and other instruments before 
reaching the CZTI detectors. The last two columns of 
Table \ref{table1} show the polar 
($\theta$) and the azimuthal ($\phi$) angles of detection of the bursts in the CZTI 
coordinates and the spacecraft components which the GRB photons pass 
through. GRB photons can scatter off the material in various parts of the satellite 
structure, affecting the detected polarization properties. It is important to 
account for these effects by means of detailed Geant4 simulations of the full
satellite. This will be discussed in later sections.

\begin{figure}
\centering
\includegraphics[width=7.4cm,height=5.2cm]{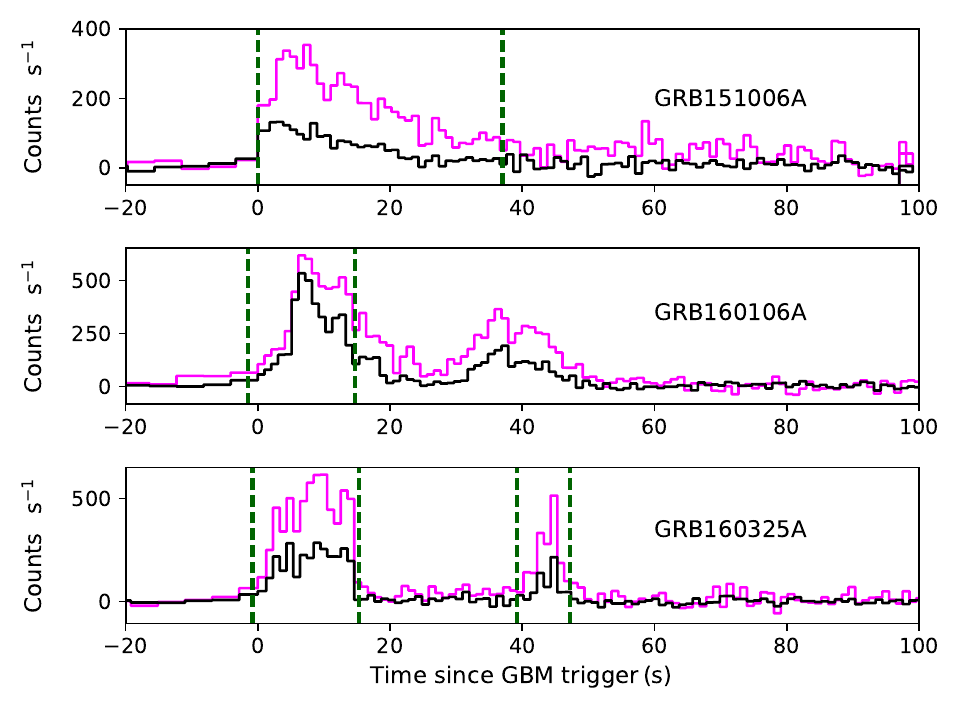}
\includegraphics[width=7.4cm,height=5.2cm]{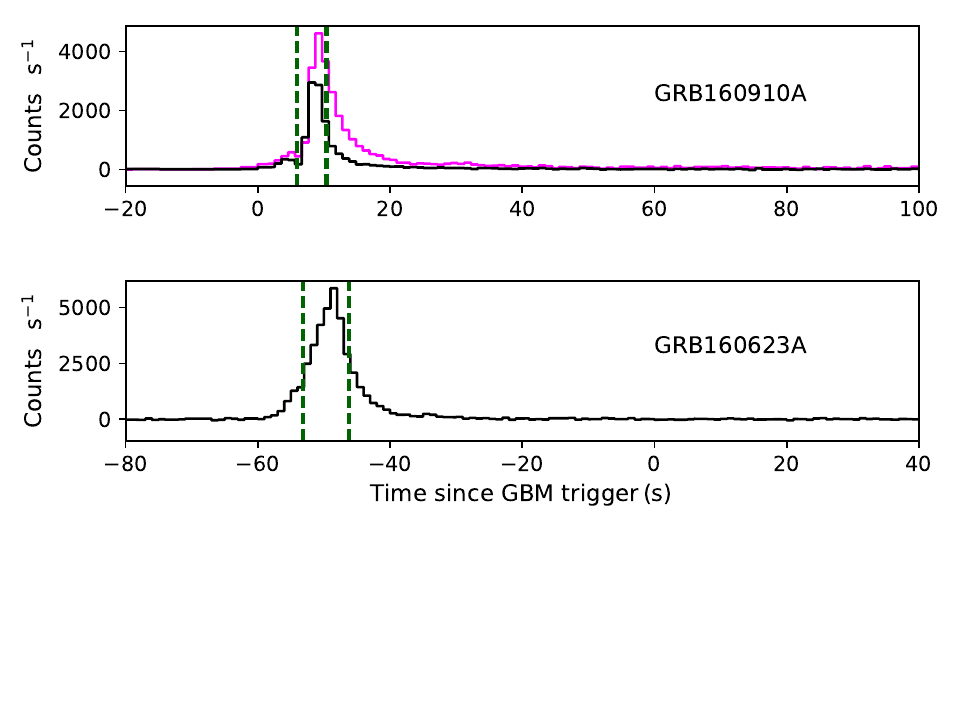}

\includegraphics[width=7.4cm,height=5.2cm]{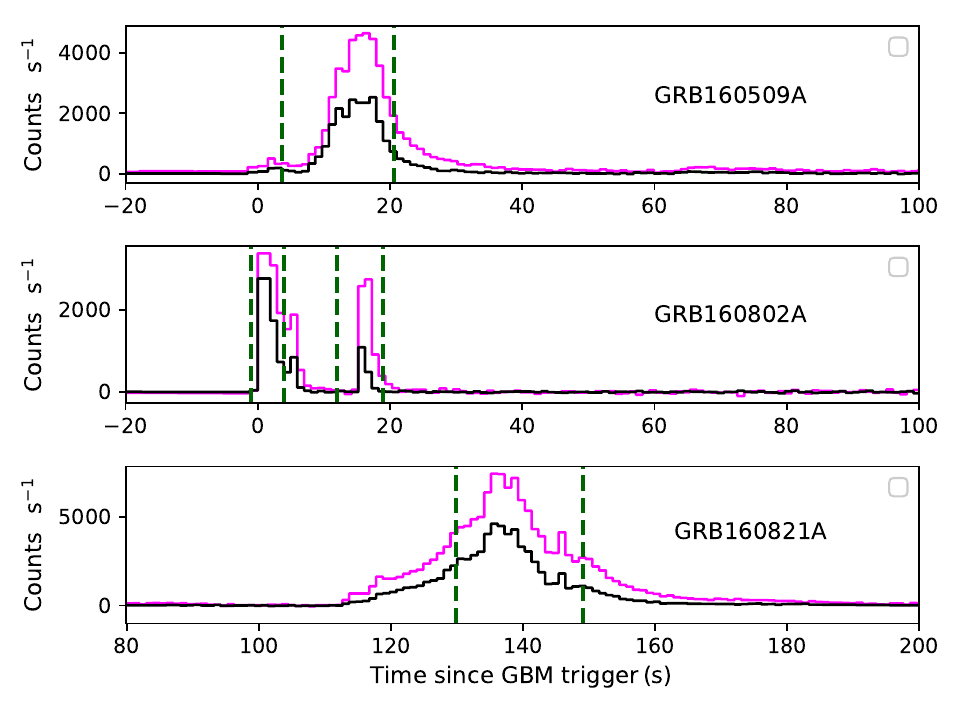}
\includegraphics[width=7.4cm,height=5.2cm]{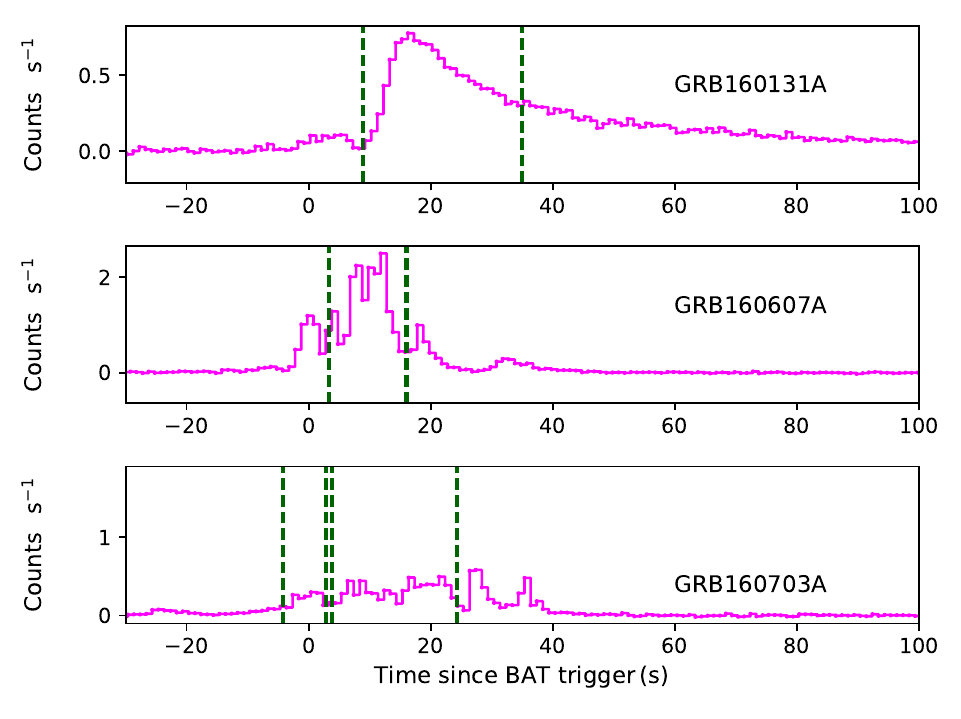}
\caption{The GRB light-curves are shown here for energy ranges 15 $-$ 100 keV 
(magenta) and 100 $-$ 300 keV (black) (see text for details). 
The vertical green lines represent 
the time intervals that have been used to extract double events for 
polarization measurements.}
\label{fig_cont1}
\end{figure}

The light curves of these 11 GRBs are displayed in figure \ref{fig_cont1}. 
The GBM light curves in 15--100 keV and 100--300 keV bands are
shown in magenta and black for 7 GRBs, while 15--100 keV BAT light curves are 
shown in magenta for the 3 GRBs detected only by {\em SWIFT}.
For GRB 160623A, the CZTI-Veto light curve in 100--300 keV band is shown in black. 
The time durations selected for polarization analysis are shown between vertical lines. 

\subsection{AstroSat mass model}
Polarization analysis of off-axis sources is challenging as the
polarization properties of photons is affected due
to the interactions with satellite elements and CZTI
housing elements. 
These interactions are highly direction and energy dependent.
To account for this we modelled the entire {\em AstroSat} with accurate chemical and geometrical properties
inside Geant4 (GEometry ANd Tracking) simulation \citep{agostinelli03}
including all the payloads of 
{\em AstroSat}: SSM, UVIT, SXT, LAXPC, CZTI and the satellite bus. The mass model is essential to model the effect of the surrounding material on unpolarized and polarized radiation (see section \ref{subsec3} and \ref{subsec4}).

\begin{figure}
\centering
\includegraphics[scale=.5]{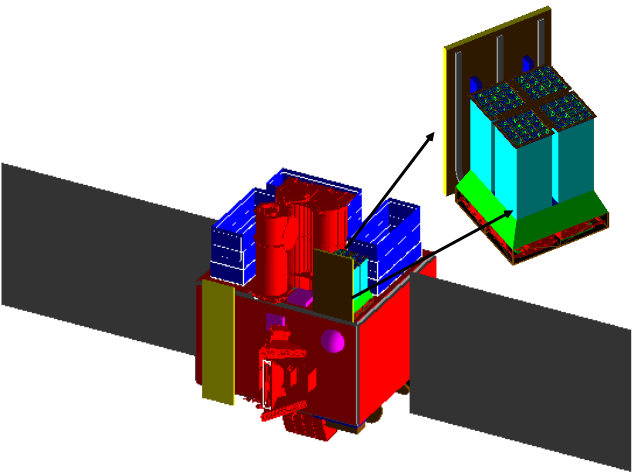}
\caption{Mass model of {\em AstroSat} simulated in Geant4 
with zoomed in view of CZT-Imager.}
\label{fig_mass}
\end{figure}

We modeled the payload and satellite bus geometries as accurately as possible. Some elements of the geometry
are coded using the GEANT4 geometry classes while
for the complex structure we used the Cadmesh interface \citep{poole11} 
to import the CAD models
into Geant4 detector construction. Figure \ref{fig_mass} shows the 
mass model of {\em AstroSat} simulated in Geant4. 
The mass model of CZTI and the Physics codes had been extensively validated during ground 
calibration of the CZTI pixels and polarization experiments with ON-axis 
calibration sources \citep{vadawale15}. 
The Geant4 geometry of other instruments (LAXPC, SXT, UVIT, SSM) and spacecraft 
were included at a later date after the launch of {\em AstroSat}. However, these 
geometries are based on the actual CAD models and hence are expected to be highly
accurate. %However, the details of
%rest of the spacecraft structure was included at later date.

Nevertheless, it is important to validate the complete {\em AstroSat} mass model by
means of observed data.
We validated the mass model in three different ways using a large number of GRBs,
including the 11 GRBs reported in the present work, which more or less
uniformly span all possible incident angles in the spacecraft reference frame. 
\begin{itemize}
    \item {\em Count distribution:} We simulated the count distribution in all 64 CZT modules using
Geant4 simulations to generate the simulated detector plane histograms
(DPH) and compared them with the observed DPH. We found that the observed 
DPH agree well with the simulated ones.
Figure \ref{dph_com} shows such distribution for GRB 160521B. The mass model distribution accounting for all on-board instruments shows a marked improvement compared to the ray-tracing results for CZTI alone.
\begin{figure}
\centering
\includegraphics[scale=.9]{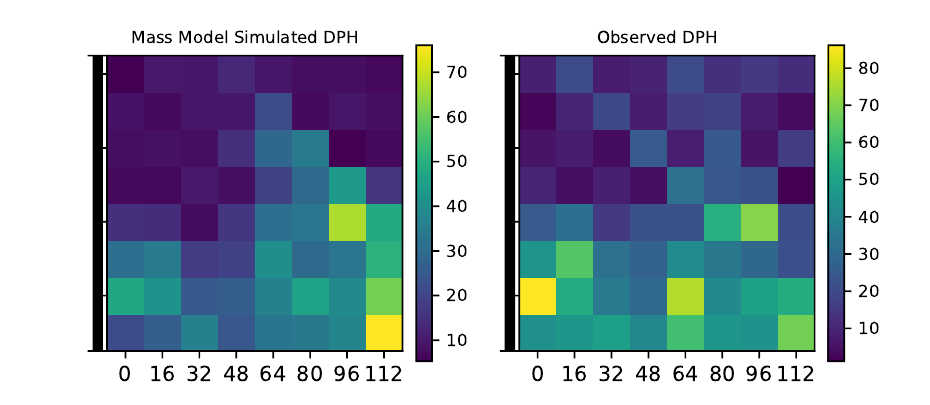}
\includegraphics[scale=.5]{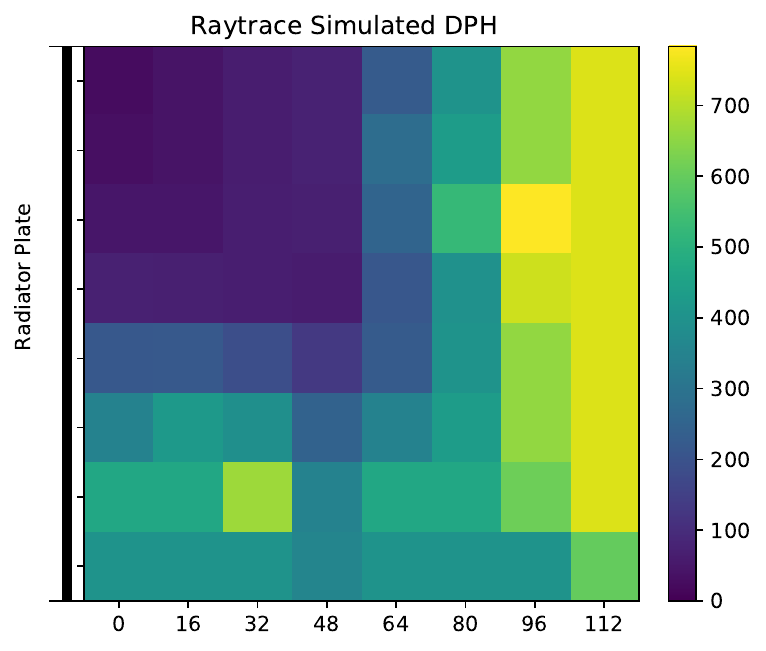}
\caption{Comparison of count distribution for GRB 160521B from mass model simulation (left top), data (right top)and ray-trace (bottom). The mass model simulated DPH agrees well with observed DPH. We also see a clear improvement in the count distribution from full satellite mass model compared to the ray-tracing distribution for CZTI alone.}
\label{dph_com}
\end{figure}

\item {\em Localization:} The DPH comparison is sufficiently good to be used for estimating the location of a GRB
for which an independent position measurement is not available. We utilized the mass model for localizing the known GRBs by DPH comparison and thereby confirming its robustness. Results for GRB 160802A is given in Figure \ref{localization}. The DPH comparison roughly peaks at the GRB location (shown as a plus sign in the figure) and it is well within the 90\% contour.
\begin{figure}
\centering
\includegraphics[scale=1.5]{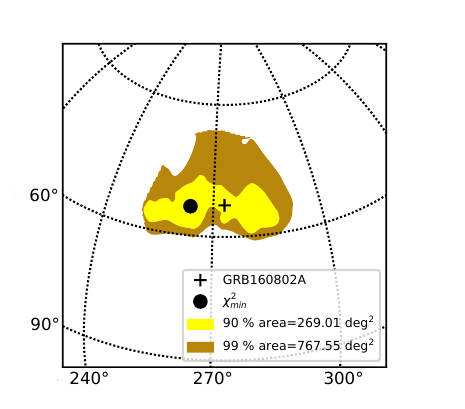}
\caption{Localization of GRB 160802A (with known localization) using DPH comparison from mass model simulation and observed data. The plus sign is the actual GRB location and the filled circle represents the peak of DPH comparison.}
\label{localization}
\end{figure}

\item {\em Broadband spectroscopy:} Using Geant4 simulations of the mass model, we generated the response for X-rays impinging in the direction of the GRB. Responses were generated for both single pixel events and the 2-pixel Compton events. We simulated mono-energetic incident photons from 50 keV to 1 MeV at every 20 keV and 1 MeV to 2 MeV at every 200 keV and record the energy distribution for each of the mono-energetic inputs. The distribution was then convoluted with a Gaussian of appropriate width. For Compton response, the total absorbed energy for a given incident photon was obtained by adding the energies in scattering and absorbing pixels. Usually, the Compton response is sensitive in the 100$-$350 keV band. However, if we utilize the low gain pixels in CZTI (around 20 \% of the detector plane), the energy response extends up to 600 keV. Although for polarization analysis, we do not use the low gain pixels, we utilized these pixels for better Compton spectroscopy.  
We extract GRB spectra from CZTI data and {\em Fermi} data for the same start and end time used in polarization analysis. The GRB spectrum for single pixel events is generated using standard method whereas the Compton spectrum is obtained by selecting the Compton events in the same GRB region and adding the energies of 2-pixel events. For background we select the pre and post GRB regions and add them.  

Figure \ref{spectra} shows the broadband fitting of GRB 160821A using {\em Fermi} and CZTI data, where the single pixel and double pixel Compton spectra from CZTI are shown in light blue and magenta respectively. The same spectral model measured using {\em Fermi} data was used here for fitting (see Table \ref{table1}). Only the relative constant for CZTI data was kept free and found to be around 0.12 $\pm$ 0.02, within error bar of the {\em Fermi} normalization of 0.14 $\pm$ 0.02. 
%The slightly lower normalization could be due to the fact that for cleaning of cosmic bunch events, we employ certain module and pixel threshold in number of events (both in onboard software and CZTI pipeline) beyond which the events are rejected. Since the GRBs are bright, we expect a fraction of source events to be lost in the process.   
\begin{figure}
\centering
\includegraphics[scale=.5]{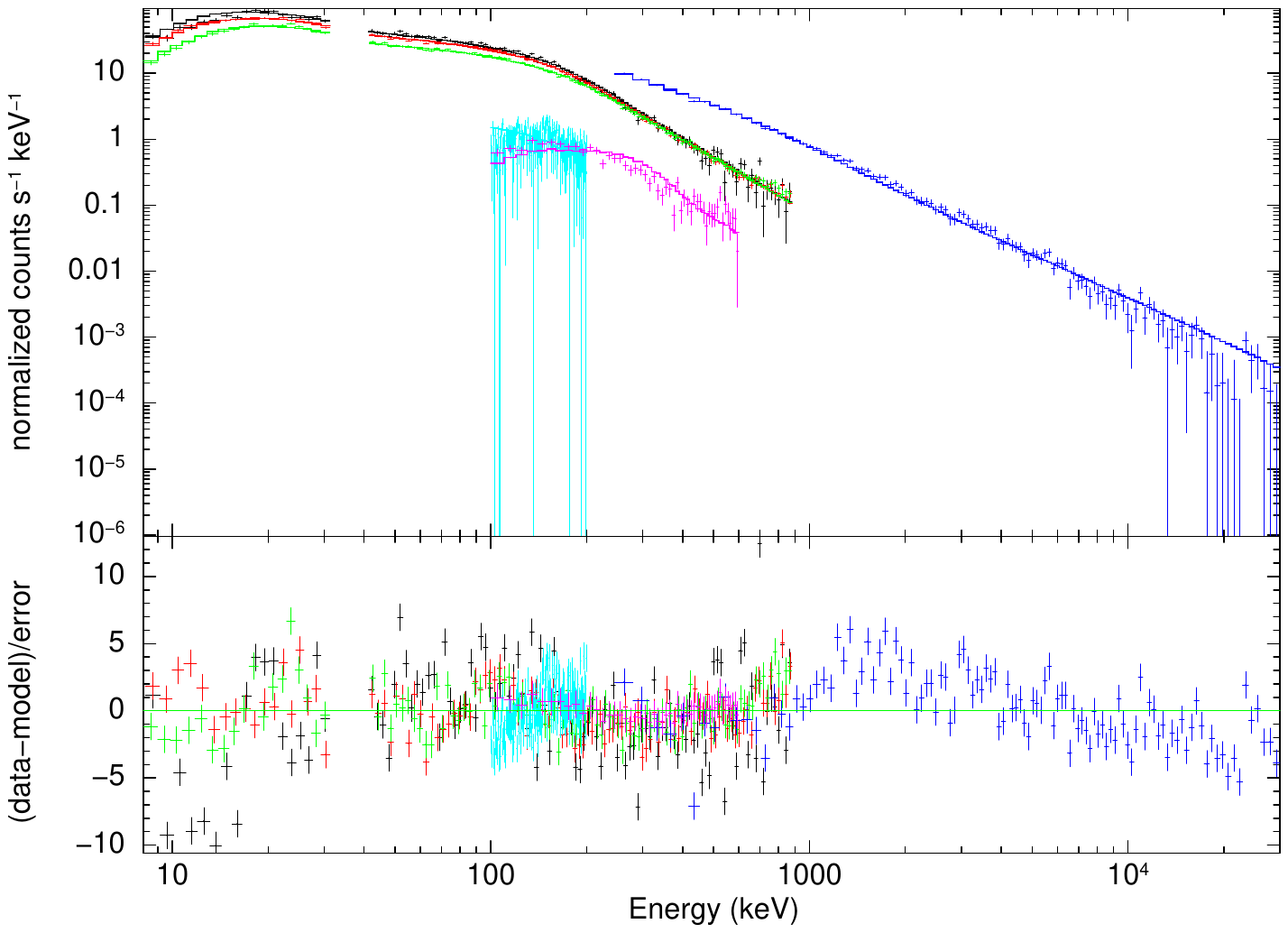}
\caption{Broadband spectroscopy of GRB 160821A with the same {\em Fermi} model given in Table \ref{table1}. The light blue and magenta are for CZTI single pixel (100-200 keV) and double pixel Compton events (100-600 keV). The black, red, green and blue are for NAI 06, NAI 07, NAT 09, and BGO 01 respectively.}
\label{spectra}
\end{figure}
The spectral shape for both the single and Compton spectra agree quite well which demonstrates the robustness of the {\em AstroSat} mass model. It is to be noted that the Compton spectrum was obtained from the same Compton events used for polarimetry. 
Further, here we stress that we have not refitted the spectrum, we used the same {\em Fermi} parameter values (except for CZTI normalization) for the combined {\em Fermi} and CZTI data to reproduce the observed spectra.   

It is to be noted that for our region of interest above 100 keV, interaction of photons with low atomic number (Z) elements is not important. Discrepancy in the low Z element geometry might get reflected in the DPH and localization which are reproduced at lower energies. At higher energies, it is the high Z element geometries which dominate the spectroscopy and polarization output. The spectroscopy results shown here clearly demonstrate that the geometries of the high Z elements have been defined accurately. This directly signifies the robustness of the polarization results to be discussed in the later sections. 
\end{itemize}

\subsection{Background event analysis}\label{subsec3}
\begin{figure}
\centering
\includegraphics[scale=.5]{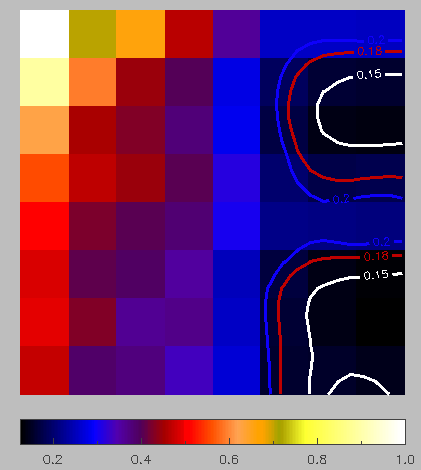}
\caption{Integrated effective area of CZT Imager (module-wise) in 
100--300 keV band for GRB 160509A. We simulate the {\em AstroSat} mass model 
in Geant4 to estimate the effective area of the modules and pixels
for the same photon energy distribution and off-axis viewing angle as for the 
observed GRB. The effective
area has been normalized with respect to its maximum value. The contours shown
in white, red and blue enclose the pixels with normalized effective area 
of 15\%, 18\% and 20\% respectively.}
\label{fig_eff}
\end{figure}

In \citet{chattopadhyay14} and \citet{vadawale15}
the various sources of background events have been discussed in detail. The
most significant contribution to background comes from the earth's
albedo radiation and diffuse cosmic X-ray background across the side 
collimators
and supporting structure, which go through Compton scattering in the CZTI 
pixels. However, we see that the background rate obtained from the onboard
data is 2--3 times higher than the values estimated from numerical simulations 
\citep{vadawale16}. 
In the uncleaned event data from CZTI, we observe cosmic ray showers in the
CZTI detectors. Though we filter out the cosmic ray events, there
is a certain probability that a fraction of these events still passes through the 
filtering conditions giving rise to a higher than expected background rate.    
The various levels of transparency of the collimators and the supporting
structures result in an unequal distribution of effective area across 
the detector pixels.
This results in a shadow pattern in the detector plane for
the observed GRBs. It is possible to suppress the background events by 
selecting the events only from the pixels with higher effective area. 
%To
%estimate the pixel-wise effective area for a given angle %of incidence, we utilized the {\em AstroSat} mass model. 
%we developed an {\em AstroSat} mass model
%inside Geant4 (GEometry ANd Tracking) simulation \citep{agostinelli03}
%including all the payloads of 
%{\em AstroSat}: UVIT, SXT, LAXPC, CZTI and the satellite. 
%Geant4 is a C++ based Monte Carlo simulation 
%toolkit which can be used to accurately simulate the passage of 
%particles through matter
%We used the Cadmesh interface \citep{poole11} 
%to import the CAD models of the payloads 
%into Geant4 detector construction. Figure \ref{fig_mass} shows the 
%mass model of {\em AstroSat} simulated in Geant4. 
%\begin{figure}
%\centering
%\includegraphics[scale=.5]{mass_model.png}
%\caption{Mass model of $AstroSat$ simulated in Geant4 
%with zoomed in view of CZT-Imager.}
%\label{fig_mass}
%\end{figure}
In order to estimate the pixel-wise effective area for a GRB, simulation is done 
for a large number (10$^9$) of incident photons in the energy range of 
100--400~keV. 
In the simulation, we employ the processes for low energy X-ray photons -- 
G4LowEnPolarizedPhotoElectric, G4LowEnPolarizedRayleigh, 
G4LowEnPolarizedCompton, G4LowEnBremss, G4LowEnIonization.  
The current simulations are done using version 4.10.03 of Geant4. 
Interaction positions, energy depositions and all 
other relevant information are stored in the output event file. 
Further analysis of selecting the valid Compton events and 
effective area estimations are done using an IDL code. 
Figure \ref{fig_eff} shows
the estimated effective area of the CZTI detector modules 
for GRB 160509A ($\theta$ = 105.7$^\circ$, $\phi$ = 85.5$^\circ$, power-law 
index = 0.75) in 100--300~keV band.
The three contours shown in the figure enclose the pixels with 
effective area of 15\% (white), 18\% (red) and 
20\% (blue) of the maximum effective area. In our analysis we select 
events only from the
pixels with effective area $>10$\% and thus filter out a 
fraction of background events resulting in a higher signal to 
background contrast. 

An important step in the background analysis is to estimate the chance 
coincidence events during the GRB that mimic true Compton events.
In Figure \ref{fig_lc}, the red light curve is obtained for double 
but non-neighboring pixels events with the other Compton criteria kept the same. 
The small peak in the light curve during the GRB 
is because of 1) chance coincidence events of the GRB photons 
within 40 $\mu$s time window in non-neighboring pixels, and 2)
the Compton scattering events between the
non-neighboring pixels. The number of Compton events between 
non-neighboring pixels can be estimated from Geant4 simulation. 
We subtract this number from the total events under the peak to estimate 
the chance events in non-neighboring pixels during the GRB. 
The estimated chance events are found to be small in number compared to
the valid Compton events ($<1-2$\%) for the brightest of the GRBs.
These numbers agree well with the theoretically computed values based on
Poisson's chance coincidence rate in a temporal window of 40 $\mu$s. 
We are particularly interested in the chance event rate in the neighboring
pixels during the GRB as these events would mimic the Compton events,
leading to a false polarization estimation. Neighboring pixel chance 
events are expected to be comparatively smaller in number compared to 
the non-neighboring pixel chance events since the number of neighboring 
two pixel combinations is $\sim$35 times
lower than non-neighboring two pixel combinations (for 256 pixels in a 
module). Consequently, we expect the chance events to be reduced by a factor
of $\sim$35 compared to the non-neighboring pixel chance events of 1--2\%,, 
which is negligible. 

\subsection{Estimation of modulation amplitude ($\mu$) and polarization angle ($\phi_0$) and their uncertainties} \label{subsec4}
\begin{figure}
\centering
\includegraphics[scale=.5]{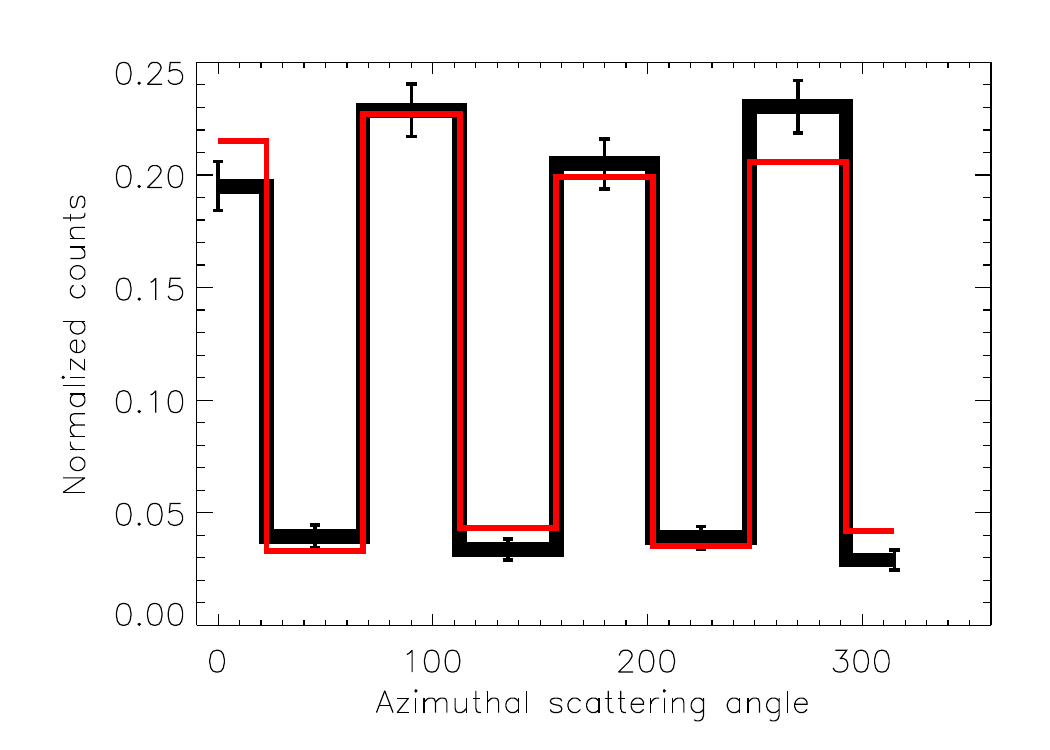}
\includegraphics[scale=.5]{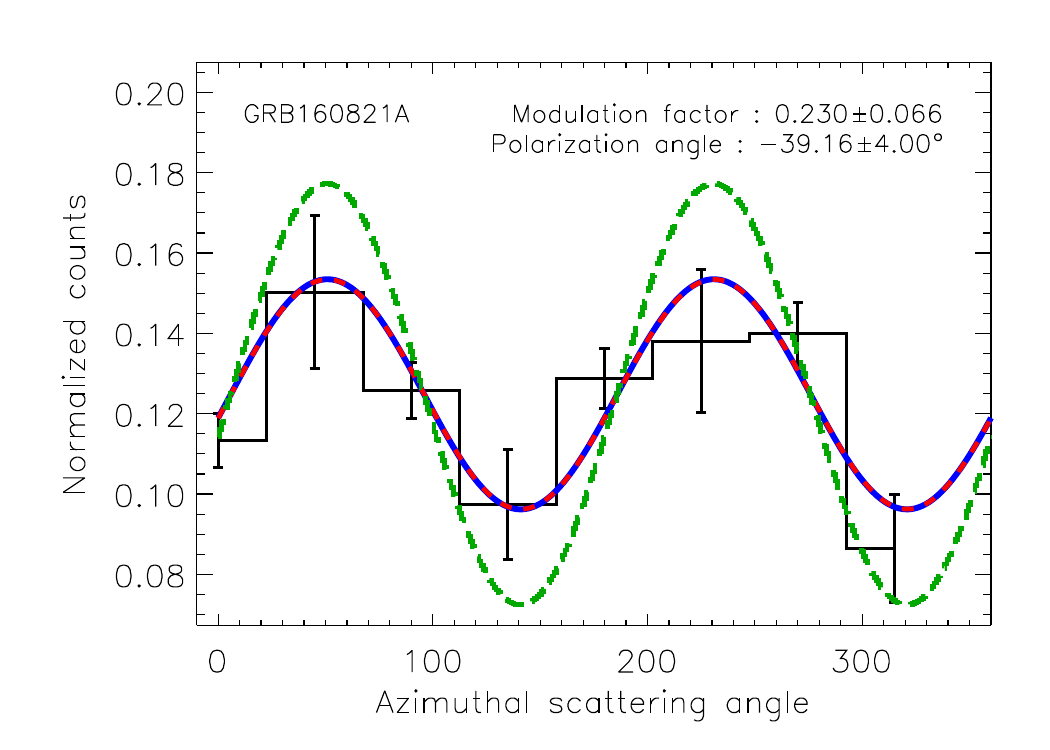}
\caption{Left: background subtracted raw eight bin azimuthal angle 
distribution for GRB 160821A obtained from the Compton 
events ($\sim$100--300 keV) are shown 
in black. The error bars are the Poisson error on each azimuthal bin 
for 68\% confidence level. The azimuthal distribution shown in red is 
that obtained by simulating unpolarized incident radiation from the 
same GRB. Right: the geometrically corrected modulation curve for 
GRB 160821A. The blue solid line is the sinusoidal fit to the modulation 
curve while the red dashed line is obtained from an MCMC method 
for a modulation amplitude $\sim$0.23 with a detection 
significance $>$3$\sigma$ (one parameter of interest at 68\% confidence level) 
and a polarization angle $\sim$-39$^\circ$ in the CZTI plane.}
\label{fig_modcurve1}
\end{figure}

In order to obtain the distribution of azimuthal scattering angles for the
GRB photons, through which the polarization signature is derived,
we first generate an 8-bin azimuthal angle distributions for combined
background and GRB events (e.g.\ the Compton events contained within the vertical 
dashed lines in Figure \ref{fig_lc}).
The azimuthal angle distribution for background events alone is then
subtracted from the total distribution to obtain the source distribution. 
The background distribution is obtained by averaging the pre-GRB and post-GRB 
azimuthal count distributions. The azimuthal angle for a given valid event
is defined with respect to the `X' axis on the CZTI plane (perpendicular 
to the radiator plate) in anti-clockwise direction when viewed from the top. 
The background count rate has been found to vary from orbit to orbit due 
to their progressively different ground traces. 
However, one of the advantages of GRB polarimetry is the availability 
of the events just before and after the GRB prompt 
emission which makes the background azimuthal distribution estimation 
comparatively easier compared to that for a persistent source.
The background subtracted azimuthal angle histogram for GRB 160821A, 
as an example, is shown in Figure \ref{fig_modcurve1} (left) in black. 
We see a significant difference in the count rate detected by the edge pixels
(angular bin $0^\circ$, $90^\circ$, $180^\circ$ and $270^\circ$) and the
corner pixels (angular bin $45^\circ$, $135^\circ$, $225^\circ$ and 
$315^\circ$). This is due to the unequal solid angles subtended
by the edge and corner pixels to the central pixel \citep{chattopadhyay14}.    It is to be noted that the azimuthal angle
distribution for any off-axis source is supposed to differ significantly
from that for an on-axis source. This is
because of the break in symmetry of the pixel geometry with respect to the 
incident photon direction. This complicates the overall shape of the 
azimuthal angle
distribution. However both these effects can be taken care of by
normalizing the azimuthal distribution of the GRB by that for a
100\% unpolarized radiation, of the same spectrum and incident at the same
off-axis angle as the source. 
If $P_i$ stands for the count of polarized photons in the $i^{th}$ angular bin, 
$U_i$ that for 
unpolarized photons in the same angular bin and $\bar{U}$ the average 
number of photons for the unpolarized distribution, then the corrected 
distribution for the polarized photon count may be obtained as
\begin{equation}
C_i = \frac{P_i}{U_i}\bar{U}.
\label{eq1}
\end{equation}
We obtain $U_i$ or the unpolarized distribution by simulating 100\% unpolarized 
incident radiation with the {\em AstroSat} mass model at the same angle of 
incidence and with the same spectrum as the observed 
GRB. The red line in Figure \ref{fig_modcurve1} (left) shows the raw 
azimuthal unpolarized
distribution, whereas the black histogram, in right panel, 
shows the modulation curve for the GRB following the correction, given in (\ref{eq1}).
 The error bars in the modulation curve
represent the 1$\sigma$ uncertainties in each bin which are mostly
dominated by the statistics of low photon counts during the GRB prompt 
emission and the uncertainty in estimating the background azimuthal
distribution. We propagate these individual contributions to finally
estimate the error in the azimuthal bins as given by Equation \ref{eq3},
\begin{equation}
\sigma_C^2(\phi)=\bigg(\frac{\bar{U}}{U_i}\bigg)^2\bigg(\frac{R_G}{T_G}+\frac{R_B}{T_G} +\frac{R_B}{T_B}\bigg)+\bigg(\frac{(C_i\bar{U})^2}{U_i^3}\bigg).
\label{eq3}
\end{equation}
The first term represents the statistical error associated with the GRB and 
background counts. $R_G$ and $R_B$ are the GRB and background count rates 
respectively. $T_G$ 
and $T_B$ are the duration of the GRB and of the selected background exposure. 
The statistical uncertainty in the background distribution can be 
made negligible with a sufficiently large background exposure. The second term in equation \ref{eq3} stands for the geometry correction (equation \ref{eq1}).
Since the geometry correction is based on Geant4 simulations of a very large number of photons (10$^9$), this term results in a negligible contribution to the final error.

In order to estimate the polarization angle and modulation amplitude 
(therefrom the polarization fraction), we use two different approaches:
1. standard fitting of the geometry corrected modulation curve by a sinusoidal
function, and 2. using Markov Chain Monte Carlo (MCMC) simulations. 

\subsubsection{Estimation by fitting of modulation curves}\label{subsubsec1}

We use the standard $\chi^2$ fitting algorithms available in IDL to fit the 
modulation curve
with a cosine function to estimate the modulation amplitude and the 
polarization angle, given by,
\begin{equation}
C(\phi) = A\, cos\,(2({\phi} - \phi_0 + \pi/2)) + B,
\label{eq2}
\end{equation}
where A, B and $\phi_0$ are the fitting parameters. The modulation
factor which is directly proportional to the polarization of the 
photons is given by the ratio of $A$ to $B$ and the polarization 
angle in the detector plane is given by $\phi_0$. 
The fitted cosine curve for GRB~160821A is shown in solid blue line in Figure \ref{fig_modcurve1} (right). The number of Compton events
used to obtain the azimuthal distribution is $\sim$2100. A clear modulation
in the azimuthal distribution signifies that the GRB is highly polarized 
with a modulation amplitude ($\mu$) around 0.229$\pm$0.062 at an angle 
-39.08$\pm$3.86$^\circ$ in the detector plane. The green dashed line is the simulated azimuthal distribution for 100 \% polarized radiation from GRB 160821A for the same observed polarization angle. The modulation amplitude ($\mu_{100}$) value is given in Table \ref{table_final}. 

\subsubsection{Estimation by MCMC simulations}
In this case the modulation amplitude, the
polarization angle and their uncertainties are estimated
using a Markov Chain Monte Carlo  \citep[MCMC,][]{geyer11} method
based on the Metropolis-Hastings algorithm \citep{hastings70,chib95}. 
The reason to follow the Bayesian statistics approach is the clarity in the 
fitting procedure and the robustness in the estimation of the parameter 
uncertainties compared to a $\chi^2$ analysis, particularly for GRBs
registering relatively fewer Compton events.  
It is not correct to assume a Gaussian distribution to estimate errors on the
polarization fraction and the polarization angle. \citet{vaillancourt06}, 
with the use of Rice distribution to compute the polarization probability
density, has shown that there is a significant departure 
from the Gaussian distribution for the low significance measurements of
polarization degree. This can be taken care of in the  MCMC simulations to 
estimate the error on polarization fraction and angle properly.
MCMC analysis also allows exploration of
Bayesian model comparison which is important to achieve a
confirmation of the detection of polarization. Therefore, we follow the MCMC 
approach for estimating the modulation 
amplitude and the polarization angle and in particular their uncertainties.
It is to be noted that, recently \citet{lowell17} used an 
advanced method of Likelihood analysis for the GRB detected by the COSI balloon 
flight \citep{chiu15}. They found better results with the new method compared to the standard $\chi^{2}$ fitting.

\begin{figure}
\centering
\includegraphics[scale=.6]{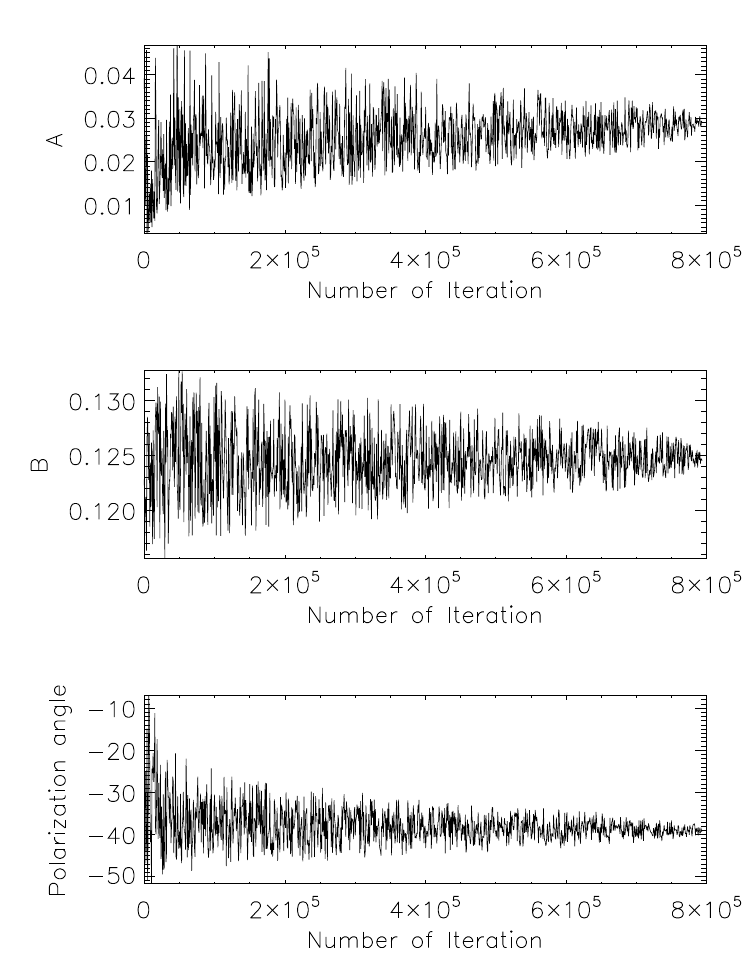}
\includegraphics[scale=.6]{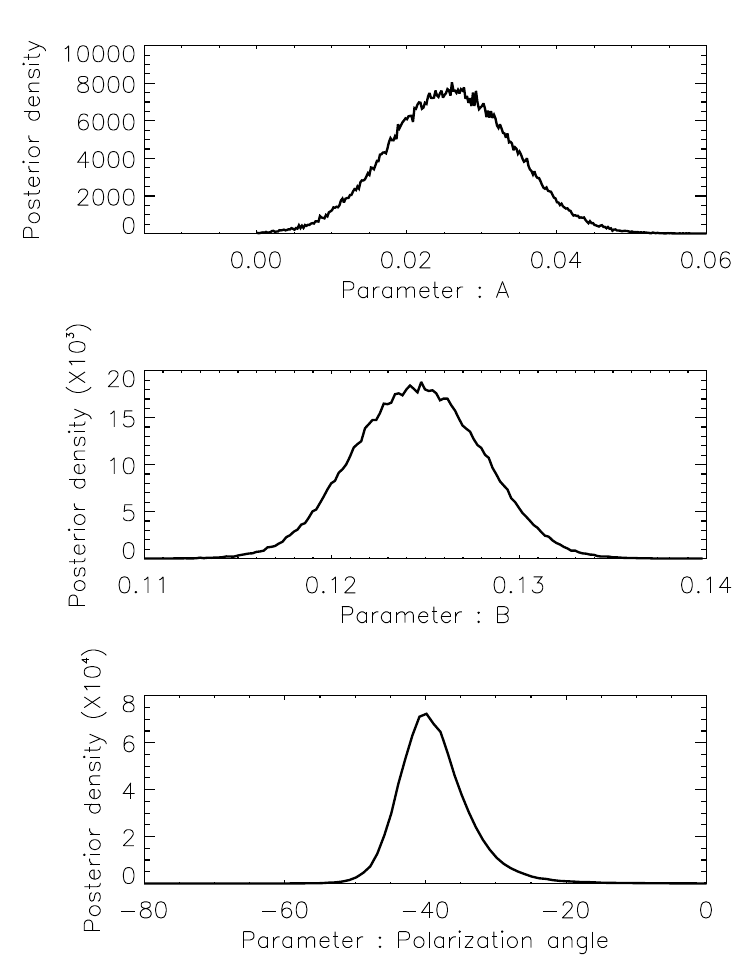}
\caption{left: Evolution of the MCMC chain with iterations for GRB 160821A. 
The MCMC simulations are done with total 1 million iterations. In the plot,
we show the evolution for intermediate 1000 interpolated iterations. Right: 
Posterior probability distribution of the fitting
parameters $A, B$ and $\phi_0$ as obtained from MCMC iterations. We compute
the uncertainties in the parameters by integrating the probability 
distribution for desired level of confidence levels.}
\label{fig_mcmc}
\end{figure}

We perform MCMC 
simulations for a large number (1 million) of iterations. For each iteration, the 
likelihood is estimated based on the randomly sampled model (\ref{eq2})
parameter values. A set of parameter values for a given iteration is accepted
or rejected by comparing the posterior probability for that iteration 
with that from the previous iteration (ratio of posterior probabilities 
should be greater than unity for accepting the parameter values). 
The posterior probabilities for those iterations with 
ratio less than unity, is further compared to a random number before
finally accepting or rejecting the parameter values. 
In this way, starting from a uniform distribution of the 
parameter guess values ($A, B, \phi_0$), we evaluate the posterior
probability for these iterations. Figure \ref{fig_mcmc} (left)
shows the evolution of the chain with iterations. While the modulation
factor and polarization angle are estimated from the best fitted 
values of the parameters ($A, B$ and $\phi_0$), uncertainties 
on them are computed from the distribution of the posterior probabilities
of the parameters. 
Figure \ref{fig_mcmc} (right) shows the posterior probability density for 
$A, B$ and $\phi_0$ for GRB 160821A. 
Uncertainties on $A, B$ and $\phi_0$ 
are estimated by integrating the probability distribution function for 
68$\%$ (1$\sigma$) confidence level. The final uncertainty on $\mu$ is 
estimated by propagating the error on $A$ and $B$. The
MCMC method yields a modulation amplitude of 0.230$\pm$0.066 and a
polarization angle -39.16$^\circ\pm$4.00$^\circ$, values that are similar to what we obtained from standard fitting.  We compared the two methods for a couple of other bright (GRB 160131A and GRB 160910A) and faint (GRB 160607A and GRB 160703A) GRBs. For GRB 160131A, we find the best fit results for polarization fraction and angle to be 0.348$\pm$0.104 and -42.7$^\circ\pm$4.89$^\circ$ respectively from the curve fitting method, while the MCMC estimates of the same are 0.347$\pm$0.116 and 41.20$^\circ\pm$5.00$^\circ$. For GRB 160910A, curve fitting gives a modulation amplitude of 0.33$\pm$0.10 and a
polarization angle -46.47$^\circ\pm$4.24$^\circ$. Similar values are also returned by the MCMC method: 0.328$\pm$0.109 and 43.54$^\circ\pm$4.00$^\circ$ respectively. We see that for bright GRBs, both the methods give
similar results, in fitted parameters as well as in their associated errors. 
For fainter GRBs, we find that the MCMC method yields slightly higher uncertainty values 
compared to the curve fitting method. For example, for GRB 160607A, the curve fitting
method results for modulation amplitude and polarization angle
are 0.209$\pm$0.203 and -42.17$^\circ\pm$11.73$^\circ$
respectively, whereas from the MCMC method the corresponding estimates are 0.206$\pm$0.2 
and -42.14$^\circ\pm$25.0$^\circ$ respectively. For the other faint burst, GRB 160703A, 
these values are found to be 0.376$\pm$0.244 and 42.94$^\circ\pm$7.38$^\circ$ from the 
curve fitting method, and 0.372$\pm$0.256 and 42.19$^\circ\pm$15.00$^\circ$ 
respectively from the MCMC method. The slightly larger errors estimated in the MCMC 
method result from the fact that MCMC explores a larger parameter space. 
Consequently, for fainter bursts, where the modulation amplitude and polarization 
angles are not strongly constrained, MCMC returns larger uncertainties.

For all the GRBs, we repeat the same procedure outlined above: namely to first 
filter the Compton events and then generate the raw azimuthal distribution, 
followed by the correction for pixel geometry and off-axis viewing angles of the GRBs. 
The corrected modulation curves are then fitted using the MCMC method to 
estimate the modulation amplitude, the polarization angle
and the associated uncertainties. The next step is to obtain the polarization
fractions of the GRBs. 
Estimation of the polarization fraction requires measurement of modulation
factor for 100$\%$ polarized radiation ($\mu_{100}$). In order to estimate
$\mu_{100}$, we simulate the {\em AstroSat} mass model in Geant4 with a large 
number of polarized photons (10$^9$)
for the same off-axis viewing angles and photon energy distribution 
of the GRBs. \citet{chattopadhyay14} show that 
$\mu_{100}$ strongly depends on the polarization angle 
and therefore it is important to estimate $\mu_{100}$ at the fitted 
polarization angles for the GRBs. 
This is done by interpolating in a table of $\mu_{100}$ values computed using Geant4 at 
a discrete grid of polarization angles.   
The uncertainty in the measured polarization angle
introduces an error in $\mu_{100}$, which is
propagated into the polarization fraction as shown in Equation \ref{eq4},
\begin{equation}
{\sigma_P}=\frac{\mu}{\mu_{100}} \sqrt{\bigg( \frac{\sigma_\mu^2}{\mu^2} + \frac{\sigma_{\mu_{100}}^2}{\mu_{100}^2} \bigg)}.
\label{eq4}
\end{equation} 
It is to be noted that the mass model simulations suggest that for 
off-axis photons the 
dependence of $\mu_{100}$ on the polarization angle is not as strong as 
in the case of on-axis photons. Apart from a few GRBs, in most cases, 
the polarization angles
have been constrained within 5--10$^\circ$ which makes this error 
negligible compared to the statistical error involved in the 
measurement of $\mu$. Details of the polarization fractions of the GRBs and
the final uncertainties will be discussed in the next section.

As discussed earlier, the results from curve fitting and MCMC methods agree well 
for the bright GRBs, while for fainter GRBs the uncertainties estimated using the
MCMC method are slightly higher. 
%This is due to the fact that MCMC explores larger parameter space in order to obtain the best fit values for $\mu$ and polarization angle. 
In order to investigate the error estimation further, we carried out Geant4 simulations
for each of these bursts for a large number of cases (10$^4$) with the same number 
of observed Compton events and used this sample to estimate the true error in 
$\mu$ and polarization angle. We then compared these error estimates with those obtained 
from MCMC. We found them to be in good agreement, with the uncertainties obtained from 
the MCMC method being slightly less than the true errors. This is because the Geant4 
estimates include the variations caused by different realizations of photon propagation 
paths through the spacecraft structures. We incorporate these additional contributions in the final estimate of uncertainties for each burst, as  described below.  

Polarization measurements are often susceptible to systematic uncertainties
and therefore it is important to take into account all possible sources
of systematics for the final error estimations. 
Here we discuss the possible systematics involved in the polarization
measurements with CZTI. 
\begin{itemize}
\item 
There can be additional uncertainty in $\mu$ due to the multiple
possibilities of interaction of the incident GRB photons with the
surrounding satellite structure. 
 As mentioned above, we estimated this through multiple ($10^4$) 
Geant4 simulations for each burst. The resulting systematic error in $\mu$ 
is found to be around 8\% for brighter GRBs (e.g. 
GRB 160821A) and around 11\% for the moderately bright GRBs (e.g. 
GRB 160131A, GRB 160802A), while it can be as high as 20\% for faint GRBs (e.g.
GRB 160703A, GRB 160607A). Additional uncertainty in the polarization angle, on
the other hand, is found to be negligible.
\item There can be systematics involved in the selection of background. 
To investigate this effect, we estimate the modulation amplitude taking both
pre and post-GRB background events independently as well as in combination. The 
estimated modulation factors and polarization angles are found to be
within $\sim$1$\%$ of each other. 
\item Polarization analysis involves normalization of the observed 
azimuthal angle distribution with respect to that for unpolarized radiation. 
The latter is obtained from Geant4 by simulating an unpolarized stream of photons 
incident at the
off-axis viewing angle of the GRB. The localization of the GRB in CZTI 
co-ordinate system is normally done based on the position provided by 
{\em Swift}/BAT or {\em Fermi}/GBM or from X-ray afterglow observations 
whenever available. The BAT position is accurate to about 3$^\prime$ 
whereas the uncertainty in GBM localizations is around 
3.7$^\circ$ \citep{connaughton15}. To investigate the effect of the 
localization uncertainty (in CZTI co-ordinates) on the estimated 
modulation amplitude, we did Geant4 simulations for 1 billion 
photons (the statistical uncertainty is negligible) using the {\em AstroSat} 
mass model in a $5^\circ\times 5^\circ$ region of the sky. 
We find the variation in modulation amplitude to be within 4\% and 
polarization angle within 2.5\%. Therefore for the GRBs localized from BAT
position, we expect this contribution to the uncertainty of modulation amplitude 
and polarization angle to be extremely small, while they can be large ($\sim 5$\%) 
for those
localized from GBM position. However, for the 11 GRBs discussed here,
the localization uncertainties are small, $<$1$^\circ$ 
(see Table \ref{table1}), contributing very little to the uncertainty in the 
derived polarization results.
\item We also investigate the dependence of the simulated azimuthal
angle distribution on the model spectra. We did mass model simulation
for GRB 160821A at the same off-axis angle but for different power-law 
spectra with index around the reported value. The dependence of the modulation 
amplitude on energy spectrum is found to be very weak with $\sim 1$\% variation 
in the azimuthal distribution.    
\item The other possible systematics in the modulation
amplitude is the unequal quantum efficiency of the CZTI pixels. However, 
since we search for GRB Compton events across the full CZTI plane, the 
relative quantum efficiency of the pixels are expected to be averaged out to a large extent.
The relative efficiency of the pixels varies only within 5\% which induces 
negligible false modulation amplitude.
\end{itemize}
Contributions from each of these sources are properly accounted for
in the final estimation of uncertainties in polarization fraction and
angle.  As an example, for GRB 160821A, the statistical error on modulation amplitude 
obtained from MCMC analysis is $\sim$0.066. With an additional 8\%
error introduced by scattering in the satellite structures, the final error on $\mu$ comes out to be $\sim$0.068 
by adding the MCMC fitting error and the additional error (8\% of $\mu$) in quadrature. 
Contributions of other sources of uncertainties (e.g. selection of background, 
localization, GRB spectra, and relative quantum efficiency of the pixels) are negligible. 
Since the error in polarization angle for GRB 160821A is very small ($\sim$4$^\circ$), 
the error in $\mu_{100}$ due to uncertainty in polarization angle turns out to be very 
small, which translates into a negligible contribution to the final error on modulation 
amplitude (see Equation $\ref{eq4}$).

\subsubsection{Calculation of Bayes factor and polarization chance
probability}\label{subsubsec2}
In spite of the significant modulations observed for the GRBs (see Figure
\ref{fig_modcurve1} and Figure \ref{fig_modcurve2}), any claim on 
polarization detection requires further investigation on the 
probability of any unpolarized radiation mimicking such modulations in the 
azimuthal angle distribution. 
This is important as the modulation amplitude is a positive definite quantity
and particularly,  we are dealing with very small number of photons 
for most of the GRBs. We adopt the Bayesian paradigm to estimate such chance 
probability.  We estimate
the Bayes factor for the sinusoidal model (for polarized photons) and a 
constant model (unpolarized photons), where the
Bayes factor is defined as the ratio of marginal likelihoods ($P(M|D)$) 
of the models: $B_{21} = \frac{P(M_1|D)}{P(M_2|D)} = \frac{P(D|M_1)}{P(D|M_2)}$,
assuming equal prior probabilities for the models ($M_1$ and $M_2$).
$P(D|M)$ or the likelihood function is computed by integrating the
posterior probability over the parameter space. There are several 
methods available in the literature for evaluating the integrals. 
We have implemented the 
\textit{`Thermodynamic Integration'} \citep{lartillot06,calderhead09}
method to compare these two models. This method allows 
the integration
in the parameter space using MCMC. We perform MCMC for each model
with $P(D|M,\theta)^\beta$ defined as the likelihood ($0< \beta < 1$), at 
different $\beta$ values and finally integrate the posterior 
probabilities over $\beta$. The Bayes Factor is eventually  
estimated from the ratio of the respective computed marginal likelihoods. 
The value of the Bayes factor, $\frac{P(D|M_{pol})}{P(D|M_{unpol})}$, required
to conclusively favor the polarized model over unpolarized radiation mimicking 
polarization signature, is
subjective and sometimes a factor $>3.2$ is considered to be substantial
proof in literature \citep{kass95}. But in our analysis we have used a
Bayes factor of 2 as the threshold for polarization measurements, namely that for 
GRBs with Bayes factor $<$2, we only estimate the upper limit of polarization. 

\begin{figure}
\centering
\includegraphics[scale=.6]{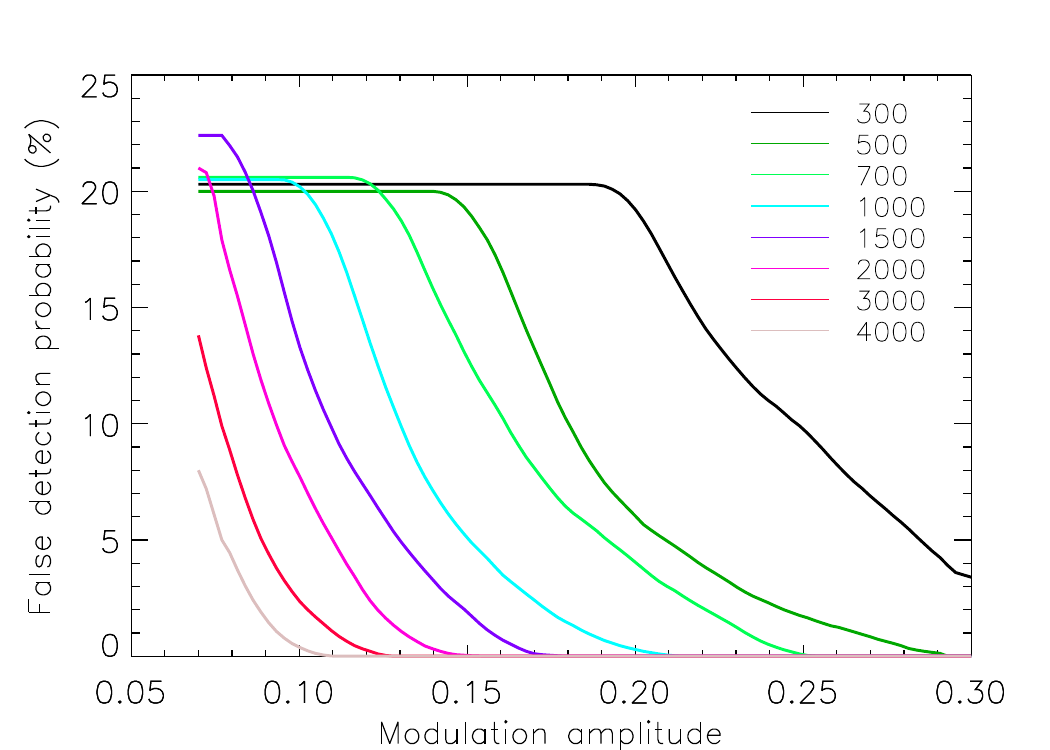}
\caption{False polarization detection probability as a function of modulation 
amplitude and the number of detected Compton events. The false probability is defined as 
the probability of unpolarized radiation resulting in a  modulation 
amplitude greater
than a reference value with Bayes factor (sinusoidal to constant fit) 
greater than 2 (see text for details).}
\label{fig_false_prob}
\end{figure}

To investigate this further, we estimate the false polarization
detection probability by simulating 100\% unpolarized radiation in
Geant4 {\em AstroSat} mass model. We repeatedly simulate unpolarized photon streams 
for a large 
number of times (10$^4$) with varying number of Compton events and estimate the 
modulation amplitude following the same method as mentioned in 
\ref{subsec4}. We define false detection probability as the  
probability of Bayes factor being $\geq$2 for estimated modulation 
amplitude equal to or greater than a given value. 
Figure \ref{fig_false_prob} shows the probability of false polarization 
detection as a function of detected modulation amplitude and the number of
Compton events.
The results shown here are obtained by simulating for off-axis 
viewing angle of GRB 160821A. We have repeated the analysis
for other viewing angles and the results are found to be similar. 
We expect the number of Compton events to be $\sim$300--4000 for 
bright GRBs in CZTI ($\sim$2100 for GRB 160821A) and therefore the 
simulations are done for Compton events in the range 300--4000. The false
detection probability is found to be as large as $\sim$20\% for 
Compton events $<$500 for detected modulation amplitude of 0.2. 
The plateau at lower modulation amplitudes (particularly for the smaller 
number of Compton events) implies that the number of false detections does
not vary below a critical value of modulation amplitude. The plateau level 
increases for Bayes factors $<$2. 
%Since GRBs are expected to be highly polarized, 
For any true polarization
detection in highly polarized GRBs, we expect modulation amplitude to be greater than 0.2. 
The number of Compton events expected for moderately bright GRBs is around
700 which makes the false detection probability very small. It is to be
noted that actual observed azimuthal angle distributions have larger 
errors due to the background subtraction. The simulated azimuthal 
distributions do not require any background subtraction and therefore 
have comparatively smaller error bars because of which the Bayes factors 
are slightly over estimated. Therefore the false detection 
probabilities obtained here represent the worst case scenario.

\subsubsection{Calculation of upper limit of polarization}\label{subsubsec3}
Upper limit on polarization are estimated for GRBs for which the
values of Bayes factor are found to be less than 2. We estimate the upper limit 
following the
method given in \citet{kashyap10}. The calculations are done in 
two steps. 
The first step involves the estimation of polarization detection 
threshold which 
we determine by limiting the probability of false detection, i.e.  
\begin{equation}
Pr(\mu > \mu_{\alpha}|P=0, N_{Compt}, N_{bkg}, BF>2) \leq \alpha,
\end{equation} 
where, $\alpha$ is the maximum allowed probability of false detection, 
$P$ is the fraction of polarization, 
$N_{Compt}$ and $N_{bkg}$ are the observed number of Compton events and 
background events for a given burst respectively and 
$BF$ is the Bayes factor, minimum value of which should be 
equal to 2 according to our chosen criteria. The false probability is 
estimated using Geant4 simulation of the {\em AstroSat} mass model for the 
observed Compton and background events for a given GRB with 100\%
unpolarized photons (as described in \ref{subsubsec2}). 
We therefore estimate the modulation amplitude, $\mu_{\alpha}$ for the 
maximum allowed probability of a false detection ($\alpha$). This is called
the $\alpha$-level detection threshold. 
In the next step, we calculate the probability of detection of polarization 
such that 
\begin{equation}
Pr(\mu > \mu_{\alpha}|P > 0, N_{Compt}, N_{bkg}, BF>2) \geq \beta, 
\end{equation}
where, $\beta$ is the minimum probability of detection. 
We simulate the GRB for the given number of source and background events 
with varying polarization fractions (from 0 to 100\%) and estimate 
$Pr(\mu > \mu_{\alpha})$ as a function of polarization fraction. The
polarimetric sensitivity of CZTI depends on the 
polarization angle in the CZTI plane. Therefore, we simulate the polarized 
photons in Geant4 at a polarization angle of 22.5$^\circ$ which corresponds to a $\mu_{100}$ averaged over 0 to 45$^\circ$ polarization angles.       
The polarization fraction ($P$) for which 
$Pr(\mu > \mu_{\alpha})$ exceeds $\beta$ gives the upper limit of 
polarization. We use values 
of $\beta = 0.5$ in conjunction with $\alpha = 0.05$ or $0.01$
for the upper limit estimations. It is to be noted that $\beta = 0.5$ actually
corresponds to the $\alpha$-level detection threshold if
we assume the sampling distribution of the estimated modulation
amplitude follows smooth Gaussian statistics with median equal to $P$
\citep{kashyap10}. A higher value of $\beta$ would correspond to 
a higher value of polarization upper limit. 
 
\section{Results} \label{sec3}
\begin{figure}
\centering
\includegraphics[scale=.35]{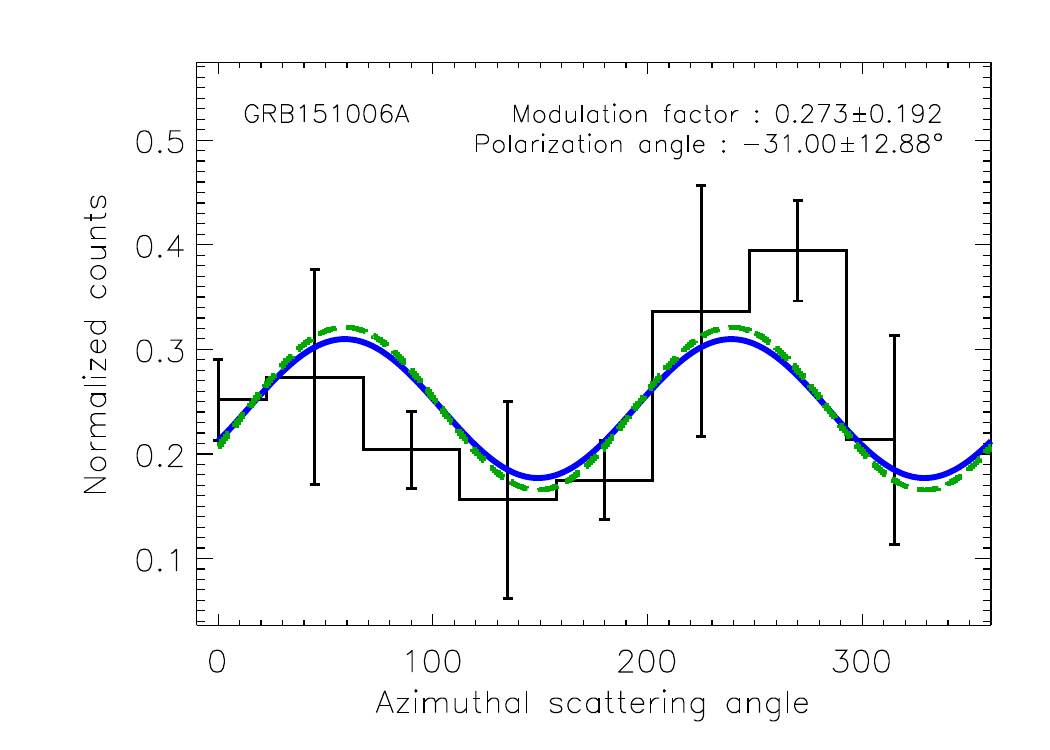}
\includegraphics[scale=.35]{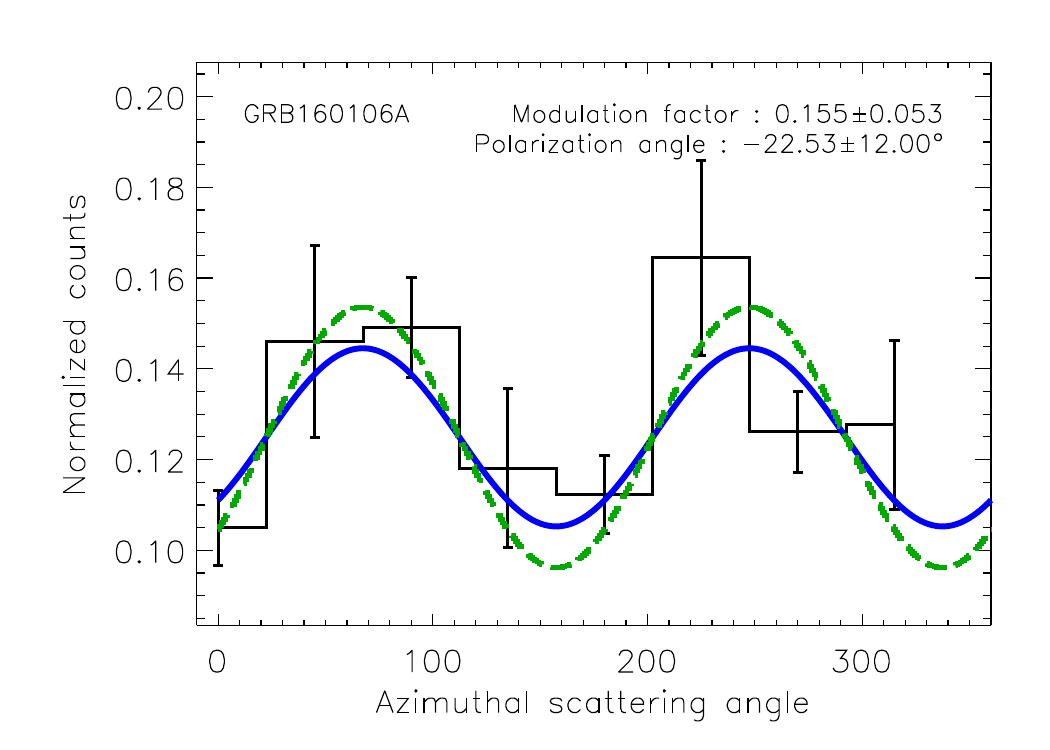}
\includegraphics[scale=.35]{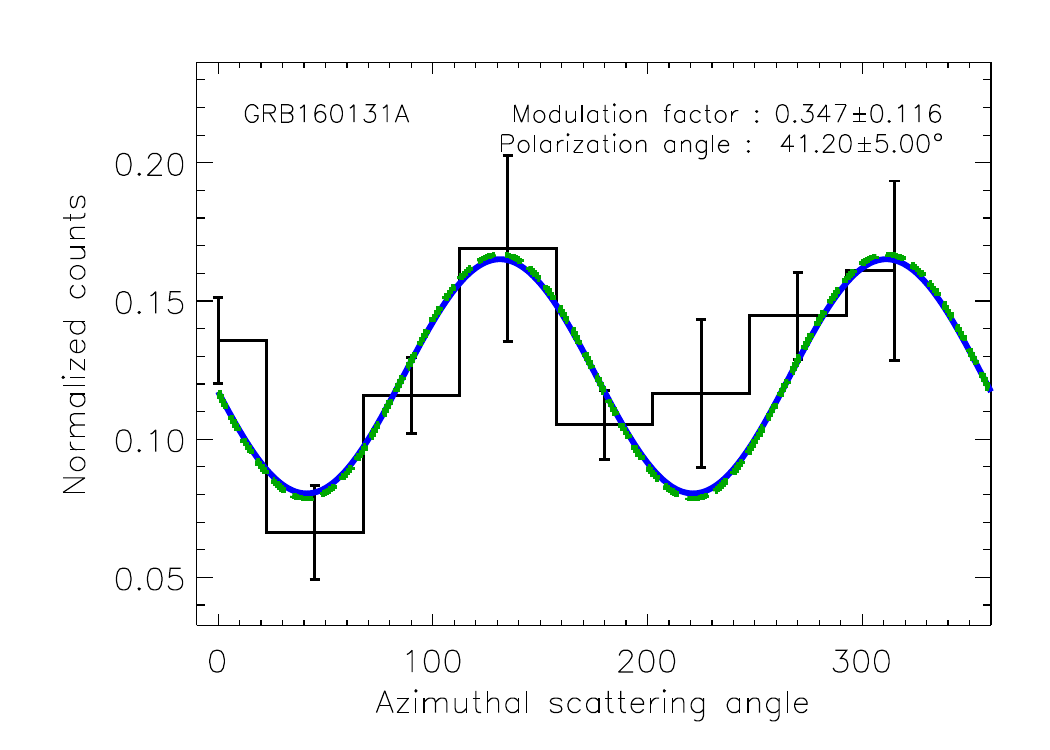}
\includegraphics[scale=.35]{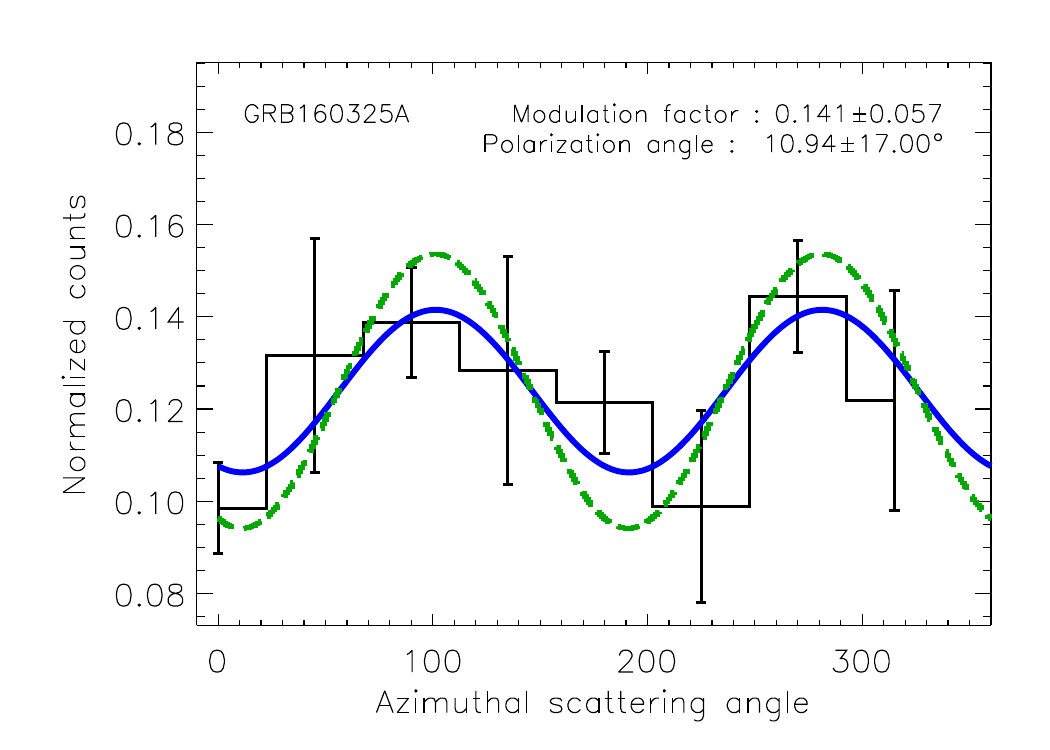}
\includegraphics[scale=.35]{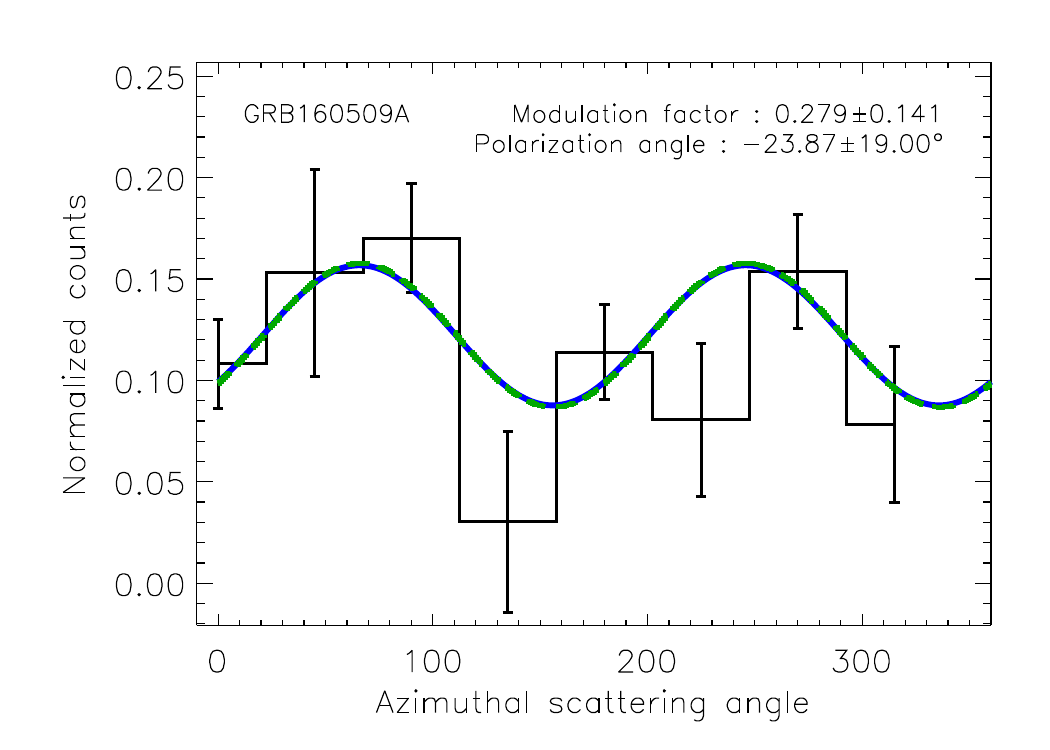}
\includegraphics[scale=.35]{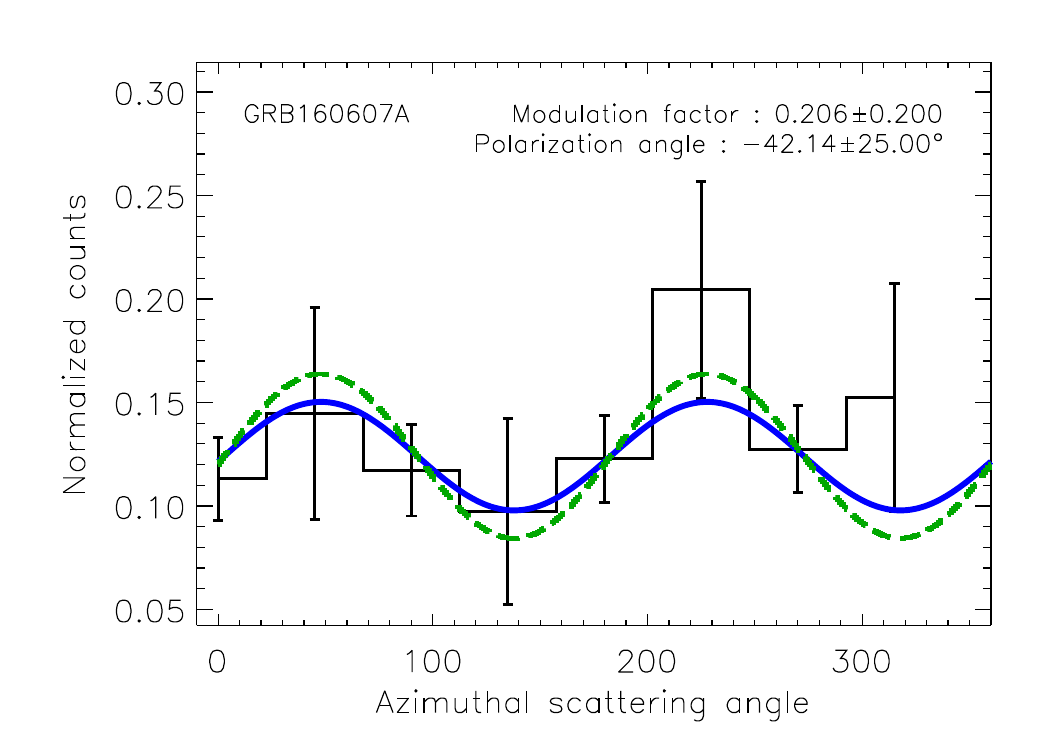}
\includegraphics[scale=.35]{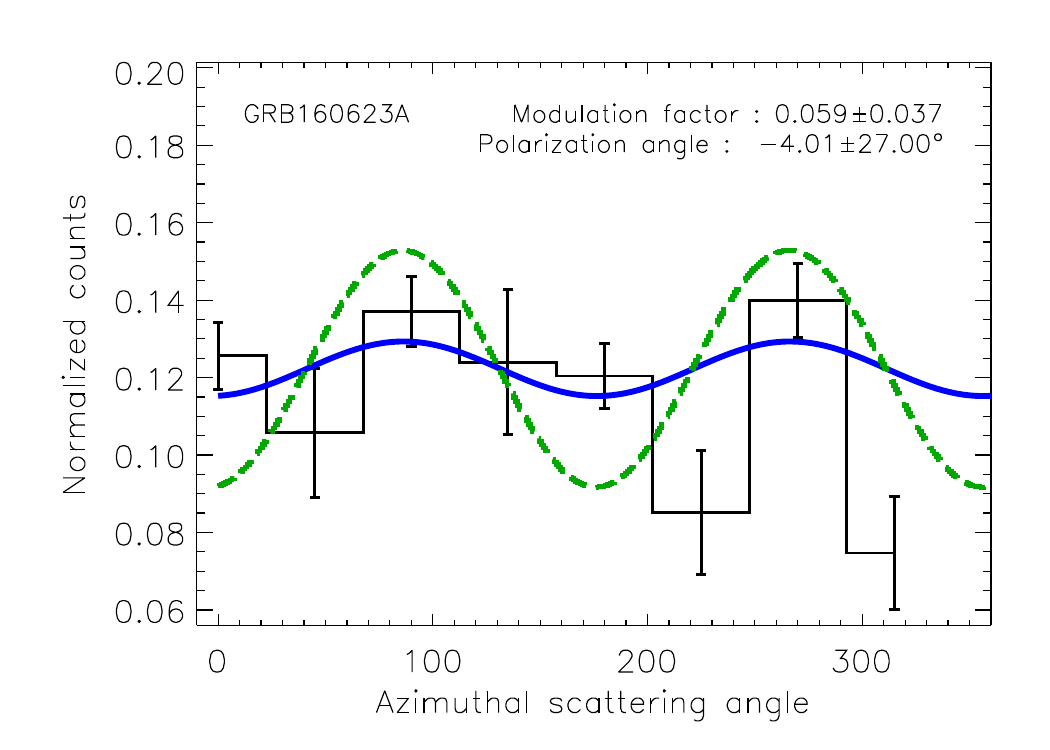}
\includegraphics[scale=.35]{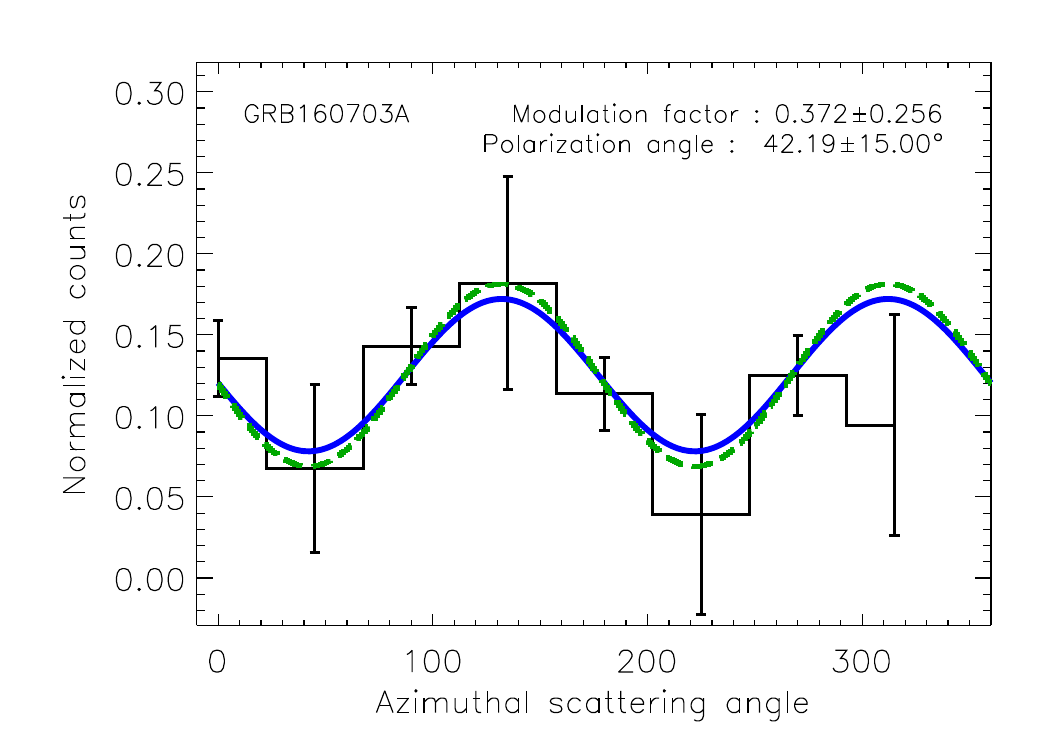}
\includegraphics[scale=.35]{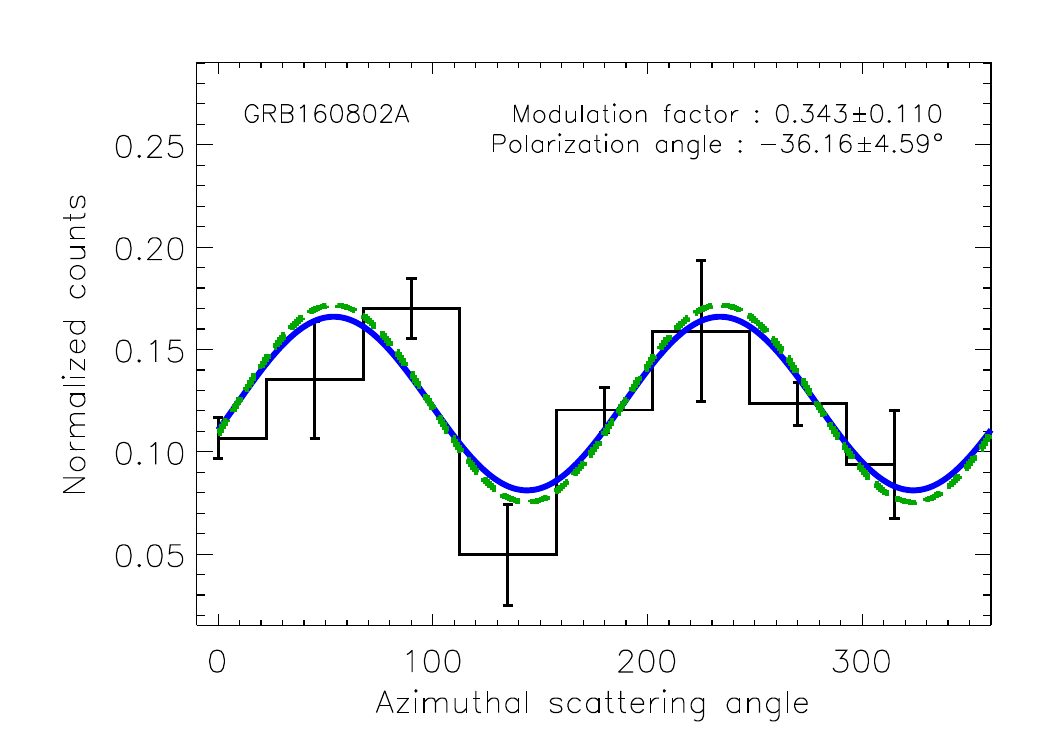}
\includegraphics[scale=.35]{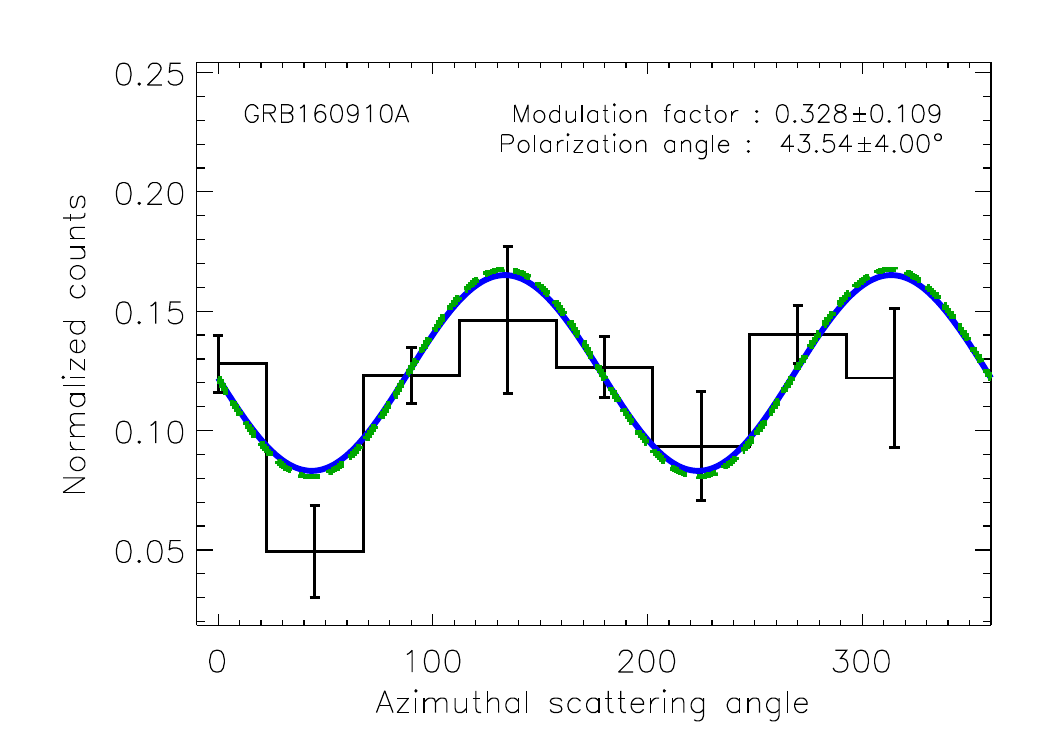}
\caption{Geometrically corrected modulation curves (similar to \ref{fig_modcurve1} right panel) for the remaining 10 GRBs. The blue solid line is the sinusoidal fit to the modulation curve while the  green dashed line is the simulated azimuthal distribution for 100 \% polarized radiation for the same observed polarization angle. Values of modulation factor and polarization angle shown in text are obtained from MCMC simulations. The uncertainties are obtained for one parameter of interest at 68 \% confidence level.}
\label{fig_modcurve2}
\end{figure}

Figure \ref{fig_modcurve2} 
shows the modulation curves for the remaining 10 GRBs. 
The modulation curves are obtained in the energy range $\sim$100--300 keV.
We see a clear polarization signature in most of the GRBs, while for a few
GRBs, lack of sufficient number of photons leads to a large uncertainty 
in the estimated modulation amplitude and the polarization angle. 
The fitted values of the modulation amplitudes and polarization angles are 
given in the text
inside the figures along with the estimated uncertainties. 
The green dashed lines are the simulated modulation for 100 \% polarized radiation for the GRBs at the observed polarization angles respectively.
Except for GRB 160325A and GRB 160802A, all the GRBs manifest a single
broad pulse. These two GRBs show two clear pulses in their lightcurves. 
The modulation curves shown here are for the combined Compton events 
from the both the peaks in order to enhance the signal to noise ratio. 
However we have seen no significant change in the modulation amplitudes and 
polarization angles across the pulses in both the GRBs. 
%with the detection significance being reduced significantly as expected.  
It is to be noted that previously we presented polarization analysis
for GRB 151006A in \citet{rao16}. The analysis was done without the 
use of detailed {\em AstroSat} mass model. With the implementation of the mass model
the new result is more accurate and the estimated modulation amplitude
is slightly less than that reported earlier.
It is to be noted that we do not see any significant modulation
for GRB 160623A in the full energy range of 100--300 keV.
The modulation amplitude is estimated to be low with large uncertainties on
both modulation amplitude and polarization angle, signifying that 
the radiation is unpolarized or has low polarization in 100--300 keV band. 
Interestingly, at energies below 200 keV, 
we find significant modulation in the azimuthal angle distribution for
GRB 160623A. It is either due to a change in the polarization angle 
or unpolarized nature of the radiation at higher energies, which leads to 
a net low polarization in the full 100--300 keV band.  Currently, it is not 
possible to distinguish these two scenarios due to poor statistics at 
higher energies. 

\begin{figure}
\centering
\includegraphics[scale=.5]{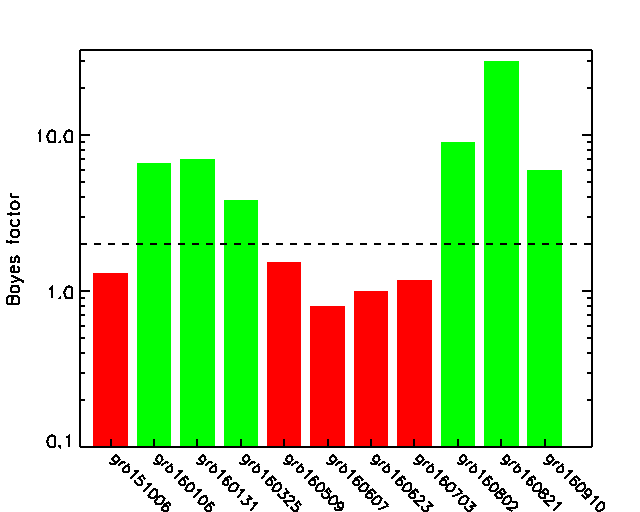}
\caption{Bayes factors for the polarized model (sinusoidal fit) to the 
unpolarized (constant fit) for all the GRBs. The Bayes factors are estimated
using combined MCMC and `Thermodynamic Integration' method (see text
for details). For bursts with Bayes factor less than 2, we estimate the 
upper limit of polarization (shown in red).}
\label{fig_bayes}
\end{figure}

\begin{figure}
\centering
\includegraphics[width=5.7cm,height=4.2cm]{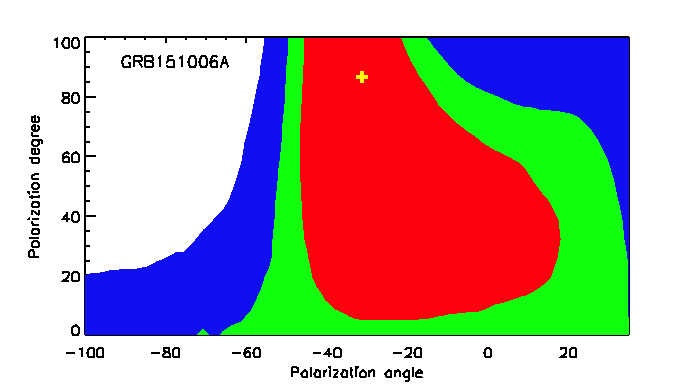}
\includegraphics[width=5.7cm,height=4.2cm]{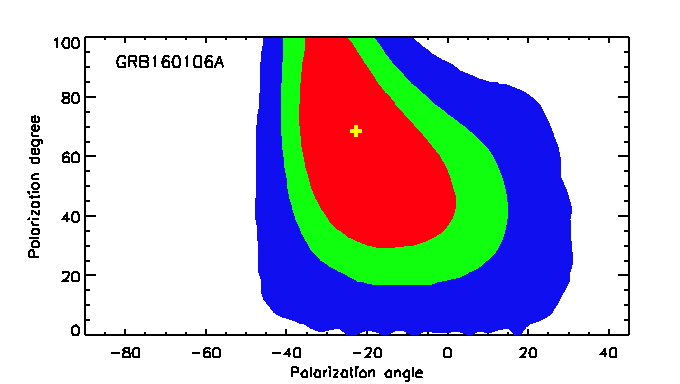}
\includegraphics[width=5.7cm,height=4.2cm]{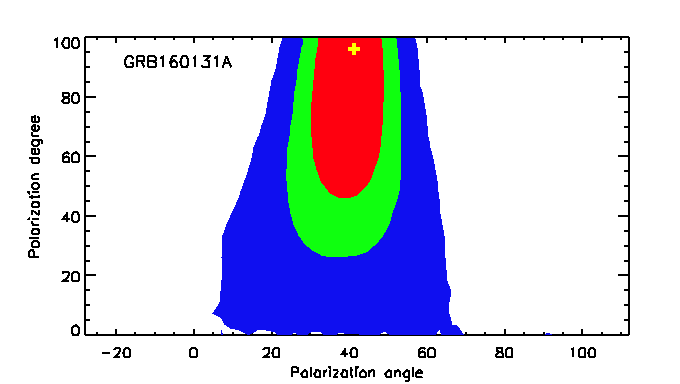}
\includegraphics[width=5.7cm,height=4.2cm]{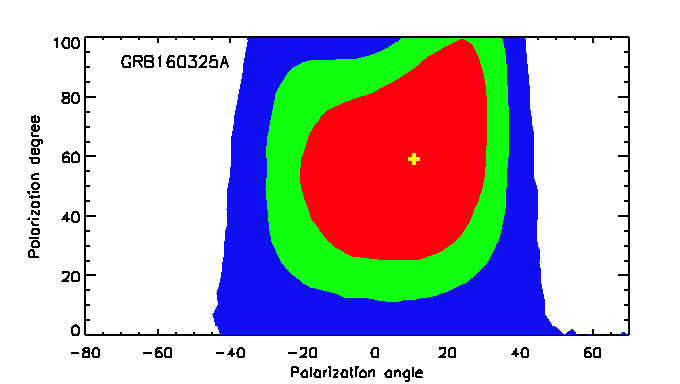}
\includegraphics[width=5.7cm,height=4.2cm]{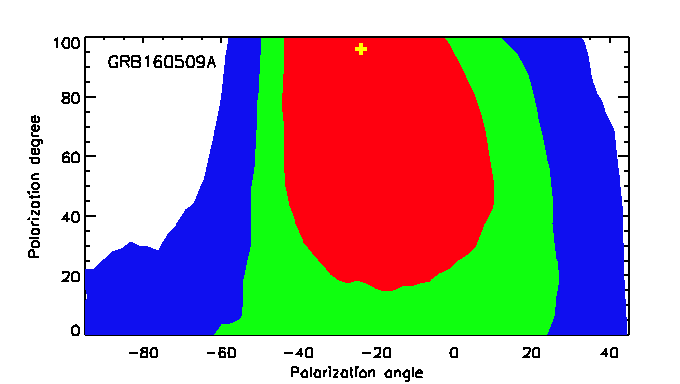}
\includegraphics[width=5.7cm,height=4.2cm]{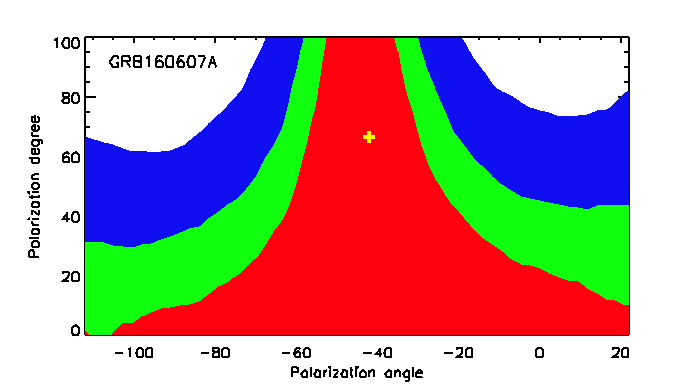}
\includegraphics[width=5.7cm,height=4.2cm]{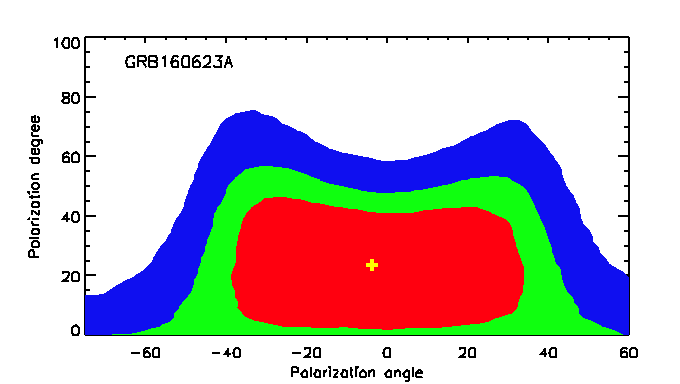}
\includegraphics[width=5.7cm,height=4.2cm]{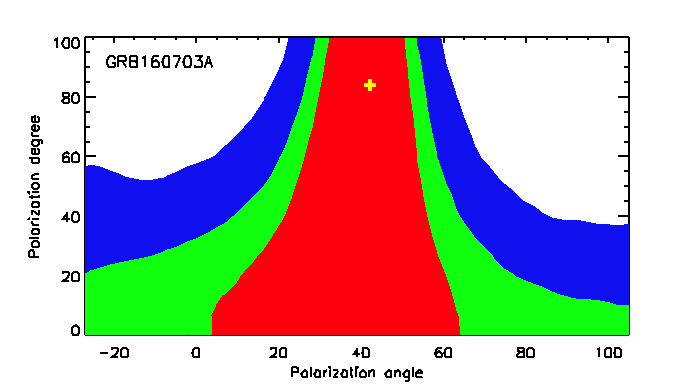}
\includegraphics[width=5.7cm,height=4.2cm]{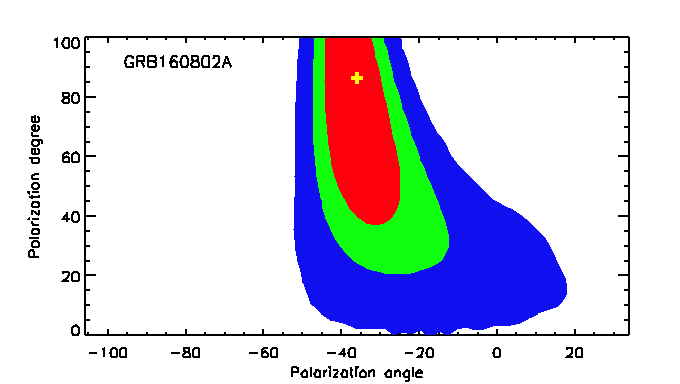}
\includegraphics[width=5.7cm,height=4.2cm]{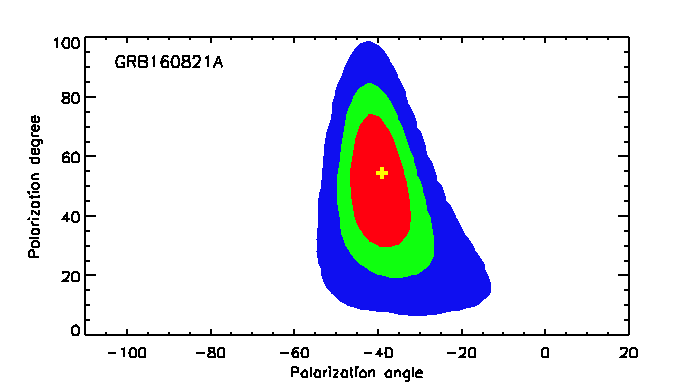}
\includegraphics[width=5.7cm,height=4.2cm]{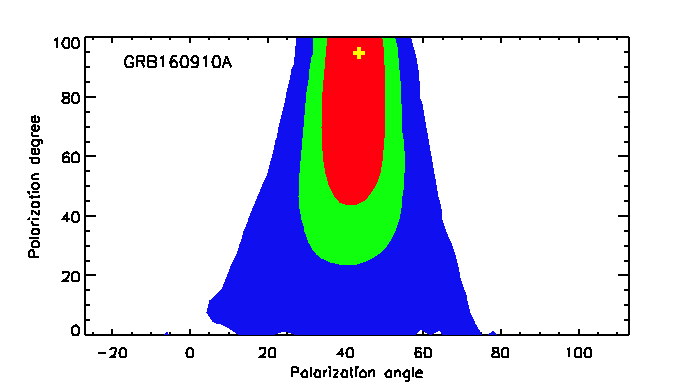}
\caption{Contour plots of polarization angle and fraction 
for all the GRBs as obtained from the MCMC method. The
red, green and blue lines represent the 68\%, 90\% and 99\%
confidence levels respectively (2 paramters of interest, polarization
fraction and polarization angle).}
\label{fig_cont}
\end{figure}

Figure \ref{fig_bayes} shows the estimated Bayes factors for the GRBs.   
We obtain high values of Bayes factor
for six GRBs, for which we can definitely claim the detection of 
polarization. GRB 151006A, GRB 160509A and GRB 160703A have Bayes factor 
slightly higher than
1, therefore the possibility that these GRBs are unpolarized can not be
completely ruled out. The probabilities of GRB 160607A and GRB 160623A being
unpolarized are high as shown in Figure \ref{fig_bayes}. For the GRBs with
Bayes factors $\leq$2, we estimate the upper limit of polarizations as 
discussed earlier.
Figure \ref{fig_cont} shows the estimated polarization fractions and the
contours for 68\% (red), 90\% (green) and 99\%
(blue) confidence levels estimated from MCMC simulations.
For GRB 151006A, GRB 160509A, GRB 160607A, GRB 160623A and GRB 160703A, 
we see that 
the polarization fractions and angles are hardly constrained. This is 
consistent with the fact that Bayes factors for these bursts are $<$ 2,
indicating that these GRBs are either intrinsically unpolarized or the degree of
polarization is below the polarimetric sensitivity of the 
instrument. 
We estimate the upper limits of polarization for these GRBs following the
method described in \ref{subsubsec3} for $\alpha=0.05$ and $0.01$ with 
$\beta=0.5$.
The derived polarization fractions and angles for the GRBs with Bayes factor $>$2 along
with the estimated uncertainties (for 1 parameter of interest with 
68\% confidence level) are given in Table \ref{table_final}. 
The uncertainties on the polarization fraction reported here are 
estimated after incorporating the systematic errors as 
discussed in \ref{subsec4}.
 Polarization fraction is estimated by normalizing the estimated
modulation amplitude with $\mu_{100}$. We estimate $\mu_{100}$ from
the Geant4 simulations of {\em AstroSat} mass model. $\mu_{100}$ depends on the
energy of the photons, polarization angle, and the incidence 
direction. \citet{chattopadhyay14} describe the dependence of $\mu_{100}$
on photon energy and polarization angle for On-axis sources. Higher values
of $\mu_{100}$ are expected when the polarization is along the corner pixels, 
whereas $\mu_{100}$ is low when it is aligned along the edge pixels. 
For off-axis angles, we find that the dependence of $\mu_{100}$ on 
polarization angle is not as significant as for On-axis sources. 
$\mu_{100}$, however, strongly depends on the incident direction 
of the photons. For larger off-axis angles, 
value of $\mu_{100}$ is found to be lower than those for smaller off-axis angles. 
In order to take these effects into account, we estimate $\mu_{100}$ by 
simulating the same GRB spectra at the same viewing angle for the 
observed polarization angle.
Values of the $\mu_{100}$s for the 
11 bursts are given in Table \ref{table_final} with the azimuthal distributions shown in Figure \ref{fig_modcurve2}.
\begin{table}
%\begin{center}
\caption{Measured polarization fractions (PF) and position angles (PA) for the GRBs}
\begin{tabular}{ c|c|c|c|c|c|c }
%CZTI GRBs
 \hline
GRB Name  & N$_{compt}$ (100 $-$ 400 keV) & PF ($\%$)$^a$ & CZTI PA ($^\circ$) & sky PA ($^\circ$) & P$_{chance}$ (\%) & $\mu_{100}^b$\\
\hline
 GRB 151006A & 459 & $<$84 ($\alpha=0.05, \beta=0.5$) & - & - & 4& 0.32 (0.27)\\
 GRB 160106A & 950 & 69$\pm$24 & -23$\pm$12$^\circ$ &108$\pm$12$^\circ$ &4& 0.23\\
 GRB 160131A & 724 &  94$\pm$33& 41$\pm$5$^\circ$ &87$\pm$5$^\circ$ & $<$0.1& 0.36\\
 GRB 160325A & 835 &  59$\pm$28& 11$\pm$17$^\circ$ &158$\pm$17$^\circ$ & 5& 0.24\\
 GRB 160509A & 460 &  $<$92 ($\alpha=0.05, \beta=0.5$)& - & - &3& 0.29 (0.24)\\
 GRB 160607A & 447 & $<$77 ($\alpha=0.05, \beta=0.5$) & - & - & 11& 0.32 (0.29)\\
 GRB 160623A & 1400 & $<$46 ($\alpha=0.05, \beta=0.5$)&  - & - & 49& 0.25 (0.29)\\
 & & $<$57 ($\alpha=0.01, \beta=0.5$) & & & \\
 GRB 160703A & 448 & $<$55 ($\alpha=0.05, \beta=0.5$)&  - & - & 0.7& 0.45 (0.41)\\
&& $<$68 ($\alpha=0.01, \beta=0.5$) & & & & \\
 GRB 160802A & 901 & 85$\pm$30 &  -36$\pm$5$^\circ$ & 147$\pm$5$^\circ$ &$<$0.1& 0.39\\
 GRB 160821A & 2100 &  54$\pm$16&  -39$\pm$4$^\circ$ & 25$\pm$4$^\circ$ &$<$0.1& 0.42\\
 GRB 160910A & 832 &  94$\pm$32&  44$\pm$4$^\circ$ & 46$\pm$4$^\circ$ &$<$0.1& 0.35\\
\hline
 \end{tabular}
 %\end{center}
\label{table_final}
\\
\\
a: $\alpha$ is the maximum allowed probability of false detection, $\beta$ is the minimum probability of detection of polarization.\\
b: The bracketed values are the $\mu_{100}$ values averaged over 0$-$45$^\circ$ polarization angles in CZTI plane. 
\end{table}
%The fifth column in the table shows the angles of polarization of the 
%GRBs estimated in the sky plane. 
 For upper limit estimations, we use $\mu_{100}$ values averaged over 0--45$^\circ$ polarization angles (given inside the brackets).
It is to be noted that the estimated polarization angles in CZTI plane
for the GRBs lie within -45$^\circ$ and +45$^\circ$. 
In order to make sure that this is not because of any systematic 
effect causing a preferred polarization direction in detector 
coordinates, we did multiple cross-checks by analyzing the GRBs at different 
energy ranges and time intervals. We find the results to be consistent to 
what is reported here. We also note that the
sky polarization angles (after converting the polarization angles in CZTI plane to the sky frame) as shown in the fifth column are
randomly oriented in the full angle space of 0--180$^\circ$ 
as expected for a large sample.
The sixth column shows the estimated false polarization detection 
probabilities for the bursts. False detection probabilities are estimated 
from the modulation amplitudes and Bayes factors of the bursts as described in 
\ref{subsubsec2}. False probabilities 
are found to be negligible for the GRBs which are bright and have 
significant detection of modulation. 
We see that most of the GRBs are highly polarized, corroborating
earlier reports for a few GRBs by {\em RHESSI}, {\em INTEGRAL} and 
GAP. For GRB 160106A, GRB 160131A,
GRB 160802A, GRB 160821A and GRB 160910A, the polarization fractions are 
estimated
with $\gtrapprox$3$\sigma$ detection significance (for 1 parameter of 
interest at
68\% confidence level). On the other hand for GRB 160325A, polarization 
fraction is constrained within $\sim$2.2$\sigma$
significance.  
 It is to be noted that the uncertainties quoted in Table \ref{table_final} 
are obtained at 68\% confidence level for only one parameter of interest, that is by 
looking only at the variation in the azimuthal angle distribution rather than 
the measurement of both polarization fraction and angle simultaneously.  The latter is 
resorted to while determining the contours presented in Figure \ref{fig_cont}. The 
estimated errors differ in these two methods. The chance probability
is also estimated from the variation in the azimuthal angle 
distribution for unpolarized radiation. Since upper limits are estimated 
from these chance probabilities, for certain bursts even though the polarization
fractions are unconstrained at 68\% level, we still obtain meaningful upper limits
on the degree of polarization.

\section{Discussions and conclusions}\label{sec4}
In the fireball scenario \citep{piran05,meszaros06}, 
interaction of highly relativistic material within the jet causes the 
prompt emission, whereas the interaction of the jet with the ambient 
medium leads to the afterglow phase. 
GRB prompt emission is widely believed to be of synchrotron origin
from high energy electrons in the jet \citep{meszaros93}. 
Apart from synchrotron, other possible mechanisms of such high energy
radiation are inverse Compton scattering, blackbody radiation and sometimes
a mixture of all these processes. The time integrated high polarization
observed in many GRBs (as shown in this work and the previously reported 
GRBs) so far demand the magnetic field to be uniform and
time independent \citep{nakar03,granot03,waxman03}, if the emission is of 
synchrotron origin.  Both the conditions are satisfied if we assume
a toroidal magnetic field geometry at large distances from the 
compact object where the radiation is emitted. This requires the field 
to be generated very close to the compact object and then carried by 
the wind which could be either Poynting flux dominated, converting the field 
energy to the kinetic energy of electrons \citep{lyutikov03} or dominated 
by the plasma particle density, where the particle energy is dissipated to
the energy of the electrons. High polarization ($\sim$40 $-$ 70 $\%$) can be 
achieved even from a random magnetic field generated in the shock plane 
itself, if the jet is narrow  ($\Gamma\theta_j\sim 1$, 
where $\theta_j$ is the jet opening angle and $\Gamma$ is the bulk Lorentz factor of
the jet) and viewed along the edge
\citep{medvedev07} or from jitter radiation by turbulence 
accelerated electrons in planar magnetic field as recently shown by
\citet{mao13,mao17}.  

\begin{figure}
\centering
\includegraphics[scale=.6]{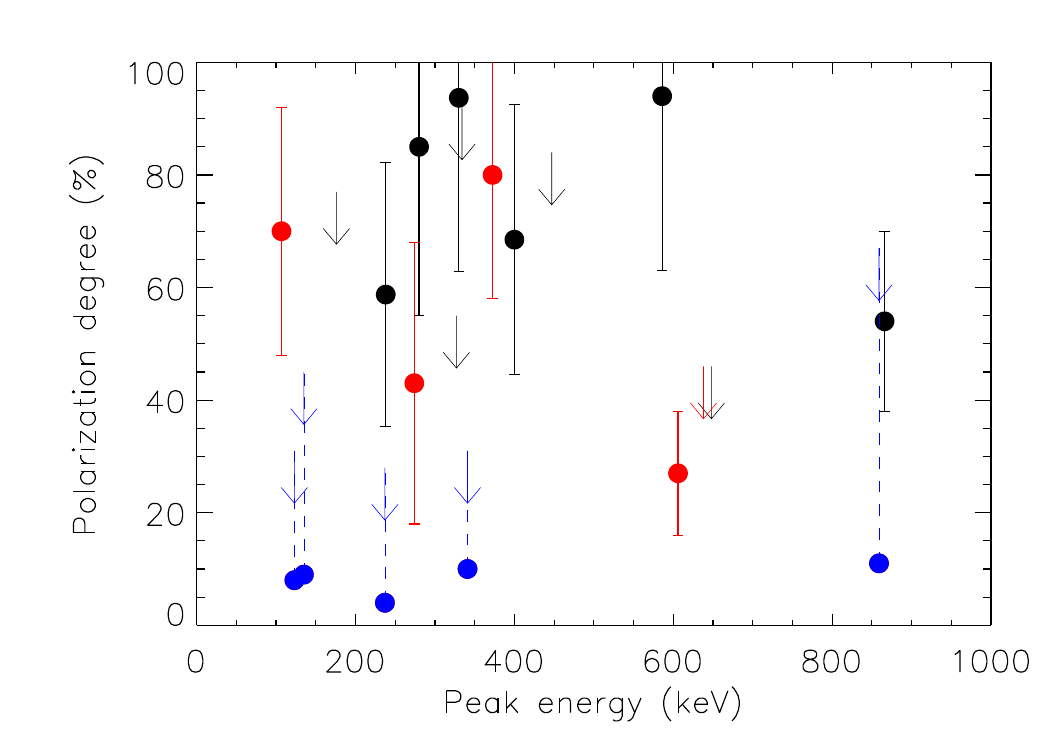}
\caption{Polarization fraction as a function of peak energies, $E_{\rm peak}$,
for the GRBs for which polarizations have been estimated. The black points
represent the GRBs detected by CZTI (see Table \ref{table_final}), while
the red points stand for those detected by GAP, {\em INTEGRAL} and COSI. The blue dots and upper limits are the polarization degree and polarization upper limits (99 \%) for the five GRBs detected by POLAR (see text for details).}
\label{fig_stat}
\end{figure}

\citet{lazzati04} demonstrated that inverse Compton emission from
relativistic electrons in a jet propagating within an external photon field 
\citep{shaviv95} can result
in high observed polarization (60--100\%) if the jet is narrow and
is observed along the edge similar to the case of random magnetic field. 
The possibility of such a geometric configuration
favorable for high polarization is relatively small for both the Compton drag
model and the random synchrotron radiation model. GRBs without such 
favorable viewing geometry are expected to be unpolarized according to 
the geometric models.
\citet{gill18} discuss these proposed radiation mechanisms and the expected levels of linear polarization for various jet geometry, viewing angle, magnetic field structure, and the spectral parameters. 
In our analysis, GRB 160623A is found to have low modulation with large errors 
in both polarization fraction and angle suggesting that the burst has
low or no polarization.  
On the other hand, we see most of the GRBs are nominally 
highly polarized ($>$70\%) 
which apparently favors the Compton drag (CD) model. This can be further
verified by investigating the $\Gamma\theta_j$ and view angle condition for Compton
drag model. 
We try to test the $\Gamma\theta_j$ condition for the GRBs with 
known redshifts (GRB 160131A, GRB 160509A and GRB 160623A). 
The isotropic energy $E_{\gamma, {\rm iso}}$ in the 
$\gamma$-ray band is found
by integrating the time-integrated spectra over 1 keV--10 MeV energy range. 
The cosmological parameters chosen were $\Omega_\lambda = 0.73$, 
$\Omega_m = 0.27$ and $H_0 ~= ~70~ km~Mpc^{-1}~sec^{-1}$ \citep{komatsu09}. 
We can find bulk Lorentz factors of the jetted emission
from the prompt and the afterglow properties by several methods 
\citep{wang17}. We relied here on the $E_{iso}-\Gamma_0$ 
correlation \citep{liang10}. The Lorentz factor decays during 
the afterglow phase, so we derived it from a prompt emission correlation. 
The initial Lorentz factor $\Gamma_0$ is constrained from the limits 
in the normalization and slope of the correlation. From $\Gamma_0$ we 
found the beaming angle of the emission ($\sim1/\Gamma_0$). 
The jet half opening angles are calculated from the jet-breaks observed 
in $Swift$/XRT X-ray 
light-curves\footnote{\url{http://www.swift.ac.uk/xrt_curves/}} \citep{sari99,frail01}. The limits on $\theta_j$
are set by selecting the radiation efficiency and the circum-burst 
density ($\eta$, $n$ ($cm^{-3}$)) in the range between (0.1, 0.001) 
and (0.9, 10). We could thus find the collimation corrected
emission energy of the bursts $E_j$ $=$ $E_{\rm iso}(1-cos\theta_j)$. 
The calculated values for these GRBs are given in Table \ref{table3}. Such estimates 
for a large number of GRBs with polarization measurements would be
very useful for a detailed understanding of the GRB prompt emission.
\begin{table*}%
%\tiny
\begin{center}
%\begin{small}

\caption{Parameters of GRB-jets derived from observed prompt and afterglow properties.}

%\hspace{-0.85in}
\begin{tabular}{cp{1.5 cm}p{1.5 cm}ccccccc}
\\
\hline\\[0.03cm]
GRB     & Redshift (z) & $E_{\gamma,iso,52}$   &  Lorentz factor ($\Gamma_0$)    &      Jet opening angle ($\theta_j)$ &$\theta_j\Gamma_0$ & $E_{\gamma,j,50}$ \\ %& $\zeta$ & \\ \hline
        &   &                       &                &                 &         &                    \\
        &   & (ergs)                &                &  ($\arcdeg$)    &         &     (ergs)         \\\\[0.03cm] \hline\\[0.03cm]  %& $\zeta$ & \\ \hline
160131A &$0.972^a$  &   $\sim$ $40$ & $459_{-49}^{+54}$ &$3_{-1.8}^{+3}$ & 25& $60_{-5}^{+18}$ \\\\[0.03cm]
160509A & $1.17^b$ & 250&$724_{-111}^{+130}$ & $4.0_{-2.3}^{+4.3}$& 51& $63_{-51}^{+203}$  \\\\[0.03cm]
160623A &$0.367{}^{c}$ &8.5 &$311_{-19}^{+21}$ &$6_{-3}^{+4}$&32&5 \\\\[0.03cm]
 \hline
\end{tabular}
\end{center}
a: \citet{ugarte16} (GCN 18966)\\
b: \citet{tanvir16} (GCN 19419)\\
%c: Hagen et al. 2016 (GCN 19656), 
c: \citet{malesani16} (GCN 19708)\\
%BAT fluence is calculated in 15 $-$ 150 keV energy range\\
%GBM fluence is calculated in 8 $-$ 1000 keV energy range
\label{table3}
%\end{small}
\end{table*}
For the remaining GRBs with no redshift measurements, we plan to use 
the Yonetoku correlation \citep{yonetoku04} to estimate the redshifts of the 
bursts.
The Lorentz factor ($\Gamma_0$) can be calculated in the same way by 
making use of the $E_{\rm iso}-\Gamma_0$ correlation. 
We plan to estimate the jet breaks and therefrom the jet opening 
angles from the available afterglow measurements.  
However, given the uncertainties associated with these
correlations and those in the measured polarization
fractions, it is difficult to constrain any of these models 
for the bursts individually. Alternatively, statistical analysis of the
prompt emission polarization for a very large sample of GRBs is expected to 
give a better insight into
the emission mechanisms behind the prompt emission \citep{toma08}. 
In Figure \ref{fig_stat} we show the estimated polarization fractions of
the GRBs as a function of their peak energies ($E_{\rm peak}$). For the GRBs detected
only in BAT, we use the $E_{\rm peak}$ values estimated from Konus/Wind time 
integrated data (see Table \ref{table1}).  
The black data points refer to the bursts studied in this work while the 
blue points represent the measurements for GRB 161218A, GRB 170101A, GRB 170127C, GRB 170206A, GRB 170114A by POLAR \citep{zhang19}. The 99 \% upper limits are also shown for these GRBs.
The red points stand for GRB 100826A, GRB 110301A,
GRB 110721A by GAP \citep{yonetoku11,yonetoku12}, GRB 160530A by COSI \citep{lowell17}, and GRB 041219A by IBIS onboard {\em INTEGRAL} \citep{gotz09}. 
The recent POLAR results indicate that GRBs are in general weakly polarized, perhaps due to the evolution of polarization angle across the burst \citep{zhang19}. In our results, we find most of the GRBs to be highly polarized. In this respect, a bigger sample of GRB polarization measurements in future will be interesting.
However, it is to be noted that we optimize the selection of energy and GRB regions slightly in order to get best detection of polarization (see Figure \ref{fig_cont1}). 
For example, GRB 160821A shows a change in polarization angle from the initial few seconds to the peak region of the burst (selected here for polarization analysis) giving rise to a net zero polarization \citep{vidushi19}. GRB 171010A also shows a change in PA across the burst \citep{chand18b}. On the other hand, from energy resolved analysis of GRB 1710101A \citep{chand18b} and GRB 160802A \citep{chand18a}, we also note that polarization properties seem to change across the peak energy of the bursts, which yields an unpolarized distribution when the individual distributions are integrated over the energies. For all the GRBs reported here except for GRB 160607A, we note that the peak energies are beyond the CZTI sensitivity range. It is interesting to note that four out of the five GRBs in \citet{zhang19} have peak energies in the POLAR energy range of analysis (80$-$500 keV). However, any concrete conclusion requires a bigger sample of polarization measurements. Given the lifetime of {\em AstroSat} of at least five years, 
CZTI is expected to detect polarization for a large sample of GRBs ($>$60).
We also plan to include the low gain pixels in the polarization analysis (shown in Compton spectrum of GRB 160821A) in order to extend the energy range to 600 keV or more which will enable us to investigate the changes in polarization properties with peak energy in addition to the enhanced sensitivity because of slightly larger collecting area. CZTI polarimetry measurements are therefore expected to provide critical inputs to distinguish the prompt emission models. 

%Given the lifetime of {\em AstroSat} of at least five years, 
%CZTI is expected to detect polarization for a large sample of GRBs ($>$60) that can provide critical inputs to distinguish
%the prompt emission models.
%The recently launched dedicated GRB polarimeter, POLAR \citep{sun16,orsi11},
%is expected to provide sensitive polarization measurements in hard X-rays
%for a large number of GRBs in the near future. All these measurements
%from CZTI and POLAR can provide critical inputs to distinguish
%the prompt emission models.  
Though synchrotron emission is widely believed to be the dominant emission mechanism behind prompt
emission of GRBs, inverse Compton scattering process and thermal emission from expanding photosphere also appear to be important in many GRBs \citep{lundman14}. 
A potential way to distinguish these various models is by investigating the dependence of polarization on the spectral and time evolution of the GRBs.
%In this context, GRB 160623A is interesting as we obtain low polarization in the full energy range whereas the GRB shows high polarization ($\sim$60\% with Bayes factor $>$1.5) at energies below $\sim$200 keV, which indicates a change in the polarization characteristics of the source at higher energies. 
A detailed spectro-polarimetric study is currently in progress for all the GRBs reported here. The preliminary spectral analysis shows a deviation from the Band model and the need for an additional thermal blackbody to 
model the spectrum more precisely for four GRBs (160106A, 160509A, 
160802A and 160910A). The GRBs 160106A, 160509A and 160910A are peculiar as the required blackbody component attains a temperature higher than peak energy ($E_{\rm peak}$) of these GRBs. We have 9 GRBs with afterglow observations and 7 of these have both optical and X-ray afterglows. Among them, 5 GRBs also have radio afterglows. A multi-band spectral and timing analysis of the prompt and afterglows emissions together with the polarization measurements can reveal more about the physics of these sources \citep{troja17}. 

\section{Summary}
% This work describes the CZTI polarimetric analysis method
% for GRBs in details and present the prompt emission polarization
% measurements for 11 bright GRBs. All the GRBs discussed here were 
% detected within the first year after the launch of {\em AstroSat}. 
% We find most of the bursts to be 
% highly polarized, implying either synchrotron emission in a time independent 
% ordered magnetic field or Compton drag as the mechanism for the  prompt emission. Given 
% the fact that all the GRBs except a couple are moderately bright, these results 
% represent the most statistically significant polarization measurements so far. 
% CZTI has almost
% doubled the number of GRBs with polarization measurements in one year and
% is expected to continue GRB polarization measurements at a similar rate for several
% years to come. Availability of a large number of such measurements from CZTI
% (along with those from POLAR) is likely to significantly enhance our understanding 
% of the GRB prompt emission. 
This work describes the  polarimetric analysis method for GRBs using the CZTI instrument of {\em AstroSat} and presents the prompt emission polarization measurements for 11 bright GRBs detected during the first year of operation of CZTI. A good polarization measurement in hard X-rays is very difficult due to two reasons:  firstly, the measurements are prone to high systematic errors and secondly, the measurement itself is of extreme photon starved nature. For the measurement of the polarization of the prompt emission of GRBs, both these aspects are significantly amplified due to the short duration of the prompt emission and the unknown position of the GRBs. These aspects are  evident from the fact that despite multiple efforts for more than a decade and a half, there has not been any firm detection of polarization apart from a few measurements made by POLAR \citep{zhang19} and GAP \citep{yonetoku11,yonetoku12}. In most cases (about 10 GRBs), only some hints of polarization have been reported (RHESSI: \citep{coburn03}, IBIS: \citep{gotz13,gotz14}, SPI: \citep{mcglynn07,kalemci07,mcglynn09}, BATSE: \citep{willis05}, {\em AstroSat}: \citep{rao16,basak17,chand18a,chand18b}, see review by \citet{mcconnell16}) and in many cases the measurements are of not very high significance. 
 
In this context, the present work is of considerable significance because it has almost doubled the number of GRBs with measured polarization in its first year of operation. Similar measurements have been carried out for a number of additional GRBs and will be reported later. It is to be noted that POLAR has stopped operating in 2017 and there are currently no GRB polarimetric mission scheduled in near future. This makes the measurements from CZTI even more important.

An important point to note here is that the results presented here (or GRB polarization measurement with CZTI in general) critically depend on the simulation for unpolarized and polarized radiation through the {\em AstroSat} satellite. For this purpose, we have made the {\em AstroSat} mass model, painstakingly collecting the details of all parts and materials gone into making the satellite. This is implemented in the Geant4 code and the resultant products (DPH, spectra, localization) are shown to agree quite well with the real data. 
%The Geant4 code for the detector configuration for Compton interaction has been validated for the On-axis polarization measurements of Crab \citep{vadawale17}. 
The residual systematics from the mass model might contribute towards the estimation of the $\mu_{100}$ but it may not have significant effect on the detection of polarization. 
%Considering the present status of GRB polarimetry, distinguishing whether a GRB is polarized or unpolarized is extremely important. 
For the GRBs with detected polarization, additional confirmation can come from comparing the polarization properties with the other observables of GRBs. We have already started this exercise \citep{chand18a,chand18b,vidushi19}.

We find most of the bursts to be highly polarized, implying either synchrotron emission in a time independent ordered magnetic field or Compton drag as the mechanism for the prompt emission. However, in order to draw such `firm' conclusions, it is necessary to have much larger sample. Given the fact that most of the GRBs in the present sample are moderately bright, CZTI is expected to continue GRB polarization measurements at a similar rate for several years to come. Availability of a large number of such measurements from CZTI is likely to significantly enhance our understanding of the GRB prompt emission.

This publication uses data from the {\em AstroSat} mission of the
Indian Space Research Organization (ISRO), archived at the
Indian Space Science Data Centre (ISSDC). CZT-Imager is built
by a consortium of Institutes across India including Tata Institute
of Fundamental Research, Mumbai, Vikram Sarabhai Space
Centre, Thiruvananthapuram, ISRO Satellite Centre, Bengaluru,
Inter University Centre for Astronomy and Astrophysics, Pune,
Physical Research Laboratory, Ahmedabad, Space Application
Centre, Ahmedabad: contributions from the vast technical team
from all these institutes are gratefully acknowledged. TC is
thankful for the helpful discussions with D. N. Burrows (PSU), 
P. Meszaros (PSU),
D. Fox (PSU), K. Frank (PSU), 
C. B. Markwardt (NASA/GSFC), V. Kashyap (Harvard) and Carson Chow (UPenn). 
This project has received funding from the European Union's Horizon 2020 research and innovation programme under the Marie Sklodowska-Curie grant agreement n. 664931.
This research has also made
use of data obtained through the High Energy Astrophysics
Science Archive Research Center Online Service, provided by
the NASA/Goddard Space Flight Center.

\bibliography{reference} % bibliography data in reference.bib
\bibliographystyle{aasjournal} 

\end{document}